%% file: main.tex
\documentclass[sort,compress]{els-mrw}

\usepackage{amsmath,amssymb,amsfonts,amsthm,makeidx,graphicx}
\usepackage{txfonts}
\usepackage{helvet}
\usepackage{slashed}
\usepackage{braket}
\usepackage{doi}

\input{definitions.tex}

\begin{document}
\chapter{Strong CP problem, theta term and QCD topological properties}

\author[1]{Claudio Bonanno}

\author[2,3]{Claudio Bonati}

\author[2,3]{Massimo D'Elia}

\address[1]{\orgname{Instituto de F\'isica Te\'orica UAM-CSIC}, \orgaddress{c/ Nicol\'as Cabrera 13-15}, Universidad Aut\'onoma de Madrid, Cantoblanco, E-28049 Madrid, Spain}
\address[2]{\orgname{Dipartimento di Fisica dell'Universit\`a di Pisa}, \orgaddress{Largo Bruno Pontecorvo 3, I-56127 Pisa, Italy}}
\address[3]{\orgname{INFN - Sezione di Pisa}, \orgaddress{Largo Bruno Pontecorvo 3, I-56127 Pisa, Italy}}

\maketitle

\begin{abstract}[Abstract]
In this chapter we introduce the $\theta$-dependence and the topological
properties of QCD, features of the strongly interacting
sector which give rise to the strong CP problem in the more general context of the
Standard Model of particle physics. We discuss the analytical approaches that
can be used to obtain qualitative, or in some cases quantitative, information
on the $\theta$-dependence of QCD and QCD-like models, discussing their range
of validity and comparing their predictions with the numerical results obtained
by means of lattice simulations.
\end{abstract}

\begin{keywords}
$\theta$-dependence \sep strong CP problem \sep topological susceptibility \sep
non-perturbative phenomena \sep lattice methods
\end{keywords}

\begin{glossary}[Nomenclature]
\begin{tabular}{@{}lp{34pc}@{}}
$\chi$PT & Chiral Perturbation Theory\\
DIGA & Dilute Instanton Gas Approximation\\
I~/~A  & Instanton~/~Anti-instanton\\
PQ  & Peccei--Quinn\\
QCD & Quantum Chromodynamics\\
WV & Witten--Veneziano \\
YM & Yang--Mills
\end{tabular}
\end{glossary}

\section*{Objectives}
\begin{itemize}
   \item In Sect.~\ref{sec2} we introduce $\theta$-dependence in a
   didactic but not completely standard way, avoiding the use of the semiclassical
   approximation; we then show how $\theta$-dependence leads to the strong CP problem.
   \item In Sect.~\ref{sec3} we describe some general features of
   $\theta$-dependence, discussing in particular the predictions that can be
   obtained by using the semiclassical approximation, the expansion in the number
   of colors, and chiral perturbation theory. These predictions are then compared with
   the numerical results obtained using the lattice formulation.
\end{itemize}


\newpage
\input{sec1_intro.tex}

\vspace*{-\baselineskip}
\input{sec2_strongCP}

\vspace*{-\baselineskip}
\input{sec3_res}

\vspace*{-\baselineskip}
\input{sec4_concl}

\begin{ack}[Acknowledgments]
We would like to thank B.~Garbrecht and F.-K.~Guo for comments.
The work of C.~Bonanno is supported by the Spanish Research Agency (Agencia Estatal de Investigaci\'on) through the grant IFT Centro de Excelencia Severo Ochoa CEX2020-001007-S and, partially, by the grant PID2021-127526NB-I00, both funded by MCIN/AEI/10.13039/501100011033. This work is also partially supported by the project “Non-perturbative aspects of fundamental interactions, in the Standard Model and beyond” funded by MUR, Progetti di Ricerca di Rilevante Interesse Nazionale (PRIN), Bando 2022, grant 2022TJFCYB (CUP I53D23001440006).
\end{ack}

\seealso{
Axions; 
Axions and Axion-like particles: collider searches; 
Chiral anomaly; 
Chiral perturbation theory; 
CP violation in the quark sector;
Effective field theory; 
Gauge topology, semiclassics and instantons; 
Large-$N$ expansion; 
Lattice field theory; 
}


\end{document}

%% file: definitions.tex
\newcommand{\beq}{\begin{eqnarray}}
\newcommand{\eeq}{\end{eqnarray}}
\newcommand{\beqnn}{\begin{eqnarray*}}
\newcommand{\eeqnn}{\end{eqnarray*}}
\newcommand{\bal}{\begin{aligned}}
\newcommand{\eal}{\end{aligned}}

\newcommand{\Tr}{\mathrm{Tr}}
\newcommand{\CP}{\mathrm{CP}}

\newcommand{\dd}{\mathrm{d}}
\newcommand{\ii}{\mathrm{i}}
\newcommand{\ee}{\mathrm{e}}

\newcommand{\subYM}{{\scriptscriptstyle{\mathrm{YM}}}}
\newcommand{\subQCD}{{\scriptscriptstyle{\mathrm{QCD}}}}

\newcommand{\supYM}{{\scriptscriptstyle{(\mathrm{YM})}}}

\newcommand{\eff}{{\scriptscriptstyle{\mathrm{eff}}}}
\newcommand{\subf}{{\scriptscriptstyle{\mathrm{f}}}}
\newcommand{\subA}{{\scriptscriptstyle{\mathrm{A}}}}
\newcommand{\subI}{{\scriptscriptstyle{\mathrm{I}}}}

\newcommand{\subE}{{\scriptscriptstyle{\mathrm{E}}}}

\newcommand{\subPQ}{{\scriptscriptstyle{\mathrm{PQ}}}}
\newcommand{\subeff}{{\scriptscriptstyle{\mathrm{eff}}}}

%% file: sec1_intro.tex
\section{Introduction}

Quantum Chromodynamics (QCD) is the non-Abelian gauge theory describing the
fundamental strong interactions among quarks and gluons within the Standard
Model of particle physics. 
While high-energy processes characterized by a large momentum transfer
can be reliably studied using perturbation theory, this is not possible in the
low energy regime, where the theory is strongly coupled. This regime is
dominated by non-perturbative effects, that are responsible for several
intriguing theoretical features which have fundamental implications for particle
physics phenomenology. The topics of this chapter---QCD topology,
$\theta$-dependence and the strong CP problem---fall under this category.

The QCD $\theta$-dependence can be seen as a
consequence of the non-trivial
topological features of the space of gluon fields, since configurations can be
classified based on the value of their \emph{topological charge} (in the following we assume the normalization $2\Tr[T_a T_b]=\delta_{ab}$):
\beq
Q = \frac{g^2}{32\pi^2} \epsilon_{\mu\nu\rho\sigma}\int \dd^4 x  \, \Tr\left[ F^{\mu\nu}(x)F^{\rho\sigma}(x)\right] \, .
\eeq
As the name suggests, this quantity has an inherently topological meaning: it
can only assume integer values for gauge configurations with finite Euclidean
action, and it is impossible to change its value via a continuous field
deformation. $\theta$-dependence is usually introduced by using semiclassical
arguments, which lead to interpret $Q$ as the number of windings of the gluon
field around the gauge group at infinite distance, but a different approach is
also possible, which will be discussed in the next section.  Being the
topological charge $Q$ dimensionless and gauge-invariant, there is no
theoretical obstruction in coupling it to the ordinary QCD action via a new
dimensionless parameter $\theta$:
\beq
\mathcal{S}_{\subQCD} \longrightarrow \mathcal{S}_{\subQCD} + \theta Q \ .
\eeq
While this topological $\theta$-term would be completely irrelevant at the
classical level (being it the integral of a four-divergence), it has actually
physical consequences in the quantum theory. Indeed the path-integral which is
expected to non-perturbatively define QCD as a Quantum Field Theory can be
decomposed into a sum over all the \emph{topological sectors} (i.~e. the
subsets of the gauge field space with fixed values of the topological charge)
and the introduction of a $\theta$-term changes the relative weight of the
different sectors, thus introducing a non-trivial dependence on this parameter
in all physical observables.

One of the most fundamental consequences of $\theta$-dependence for Standard
Model phenomenology is the explicit violation of the charge conjugation and
parity (CP) symmetry it introduces in the strong sector, being $Q$ CP-odd by
virtue of the presence of the Levi-Civita tensor. So far, however, no strong CP
violation has ever been observed in experiments, leading to a very stringent
upper bound on $|\theta|$, which is set to be of the order of $\sim
10^{-9}-10^{-10}$ by the non-observation of the neutron electric dipole moment.
This introduces a fine-tuning problem on $\theta$ known as \emph{strong CP
problem}, one of the most relevant open problems of the Standard Model.

One of the most promising proposed solution to the strong CP problem is the
Peccei--Quinn axion, an hypothetical particle whose peculiar coupling to gluons
dynamically suppresses strong CP violations. Axion detection is currently the
target of several major experimental collaborations, and has motivated plenty
of theoretical studies about axion dynamics. In this respect, accurate
predictions about QCD $\theta$-dependence turn out to be crucial to describe the
axion effective potential. While in some regimes it is possible to obtain
reliable predictions about $\theta$-dependence using suitable approximations,
one has in general to resort to numerical Monte Carlo simulations of the
lattice-discretized theory to systematically address this task from
first-principles and in a fully non-perturbative way.

The goal of this chapter is to pedagogically introduce the reader to
$\theta$-dependence in QCD and to the formulation of the strong-CP problem
(Section~\ref{sec2}), providing a compact compilation of state-of-the-art
non-perturbative results about QCD topology and $\theta$-dependence obtained by
means of both analytical tools and numerical methods (Section~\ref{sec3}).

%% file: sec2_strongCP.tex
\section{\texorpdfstring{$\theta$}{theta}-dependence and the strong CP problem}
\label{sec2}

\subsection{Why \texorpdfstring{$\theta$}{theta}-dependence}

The $\theta$-dependence of four-dimensional gauge theories is typically
introduced using the semiclassical approximation and instantons, see, e.~g.,
\cite{ColemanBook, ShifmanBook, Weinberg2Book, ZinnJustinBook}. A canonical
approach can however also be used, which emphasizes the role of locality and of
the topology of the gauge group (as opposed to the topology of field
configurations with finite Euclidean action), clarifying that $\theta$-dependence is a physical property of the theory independent of the
semiclassical approximation.  Such an approach has been developed in
\cite{Jackiw:1979ur, Jackiw85}, and its influence can be seen, e.~g., in
the presentations of \cite{WeinbergEBook} and, particularly, of \cite{SmilgaBook}.

Before addressing the case of four-dimensional gauge theories, let us start by
discussing in some detail what is arguably the technically simplest system
characterized by $\theta$-dependence, i.~e., a quantum particle moving on a
circumference.  This system, despite its extreme simplicity, displays several
key features which have analogues in four-dimensional gauge theories
\cite{Jackiw:1979ur, Rothe:1977bv, SmilgaBook, SchulmanBook, Bietenholz:1997kr,
Gaiotto:2017yup, Bonati:2017woi}.  If we denote by $\phi$ the angle on the
circumference, the Hamiltonian of a quantum particle of mass $m$ moving on a
circumference of radius $R$ is
\begin{equation}\label{eq:quantpartH}
H=\frac{1}{2mR^2}p_{\phi}^2+V(\phi)
=-\frac{\hbar^2}{2mR^2}\frac{\partial^2}{\partial\phi^2}+V(\phi)\ ,
\end{equation} 
where $p_{\phi}=-\ii\hbar\frac{\partial}{\partial\phi}$ and $V(\phi)$ is an
external field satisfying the geometrical constraint $V(2\pi)=V(0)$.  We 
moreover assume the external potential to be parity invariant, i.e.~$V(\phi)=V(2\pi-\phi)$, and real.  The Hamiltonian $H$, to represent a well
defined self-adjoint operator, needs to be supplemented by boundary conditions
on the interval $[0,2\pi)$. The periodic boundary conditions
$\psi(0)=\psi(2\pi)$ are typically imposed to the wavefunction $\psi(\phi)$,
but more general conditions can also be adopted.

A physically sensible way to find more general boundary conditions is to note
that, if we extend the angle $\phi$ from $[0,2\pi)$ to the whole $\mathbb{R}$,
the system is characterized by a $2\pi$ translation invariance in $\phi$. If we
denote by $U$ the $2\pi$ translation operator $U\psi(\phi)=\psi(\phi+2\pi)$,
this operator is thus an unitary operator which implements a symmetry of the system,
hence it commutes with the Hamiltonian $H$, and $U$ and $H$ can be diagonalized
simultaneously. Considering the eigenvalue $\ee^{\ii\theta}$ (the notation is not
accidental, as we will show in a moment) of the unitary operator $U$, the
corresponding eigenvector $|\psi_{\theta}\rangle$ satisfies
$U|\psi_{\theta}\rangle=\ee^{\ii\theta}|\psi_{\theta}\rangle$, which corresponds to
periodic up-to-a-phase boundary conditions for the wave function
\begin{equation}\label{eq:thetabc}
\psi_{\theta}(2\pi)=\ee^{\ii\theta}\psi_{\theta}(0)\ .
\end{equation}
Since $U$ commutes with the Hamiltonian, eigenspaces corresponding to different
eigenvalues of $U$ will not interfere with each other in the dynamics of the
system, and we can thus use Eq.~\eqref{eq:thetabc} as boundary condition for
the differential operator in Eq.~\eqref{eq:quantpartH}. This is just the
starting point needed to prove the Bloch theorem in solid state physics, see,
e.~g., \cite{AshcroftMerminBook}, and Eq.~\eqref{eq:thetabc} is in fact the
most general boundary condition which makes the differential operator
$p_{\phi}=-\ii\hbar\frac{\partial}{\partial \phi}$ self-adjoint on the $[0,2\pi)$
interval, see, e.~g., \cite{ReedSimon2Book}, or \cite{Bonneau:1999zq} for an
elementary discussion.  Already at this level we can see that, since  for
values of $\theta$ different from $0$ and $\pi$ we have $\ee^{\ii\theta}\neq
\ee^{-\ii\theta}$, the boundary condition in Eq.~\eqref{eq:thetabc} explicitly
breaks the symmetries under parity ($\mathrm{P}\psi(\phi)=\psi(2\pi-\phi)$) and time
reversal ($\mathrm{T}\psi(\phi)=\psi^*(\phi)$) that characterize the classical system. 

Path integration provides a nice way of moving $\theta$-dependence from the
boundary condition to the action, which will be useful also in the case of four
dimensional gauge theories. The propagator of the theory obtained without
fixing the value of $\theta$ can be explicitly written as
\begin{equation}
\langle \phi_f, t_f| \phi_i, t_i\rangle= \int_0^{2\pi} 
\dd\theta \sum_n  \psi_{\theta, n}(\phi_f)\psi_{\theta, n}^*(\phi_i) \ee^{-\ii E_n^{(\theta)} (t_f-t_i)/\hbar}\ ,
\end{equation}
where $\psi_{\theta, n}$ is the $n$-th eigenfunction of the problem with
boundary condition Eq.~\eqref{eq:thetabc}, and using path integration (see,
e.~g., \cite{SchulmanBook, FeynmanHibbsBook, ZinnJustinPIBook}) this expression
can be rewritten in the form (we always consider $\phi\in\mathbb{R}$)
\begin{equation}\label{eq:pathint}
\langle \phi_f, t_f| \phi_i, t_i\rangle= 
\int_{\phi(t_i) \, =\, \phi_i}^{\phi(t_f) \, =\, \phi_f} \mathcal{D}\phi(t) \, \exp\left(\,\frac{\ii}{\hbar}\int_{t_i}^{\,t_f}L(\phi,\dot{\phi})\dd t\right)\ ,
\quad L=p_{\phi}\dot{\phi}-H=\frac{1}{2}mR^2\dot{\phi}^2-V(\phi)\ .
\end{equation}
Since $\ee^{\ii\theta}$ is the eigenvalue of the operator $U$ which has been
diagonalized simultaneously with $H$ we also have
\begin{equation}
{}_{\theta'}\langle \phi_f, t_f| \phi_i, t_i\rangle_{\theta}= \delta(\theta-\theta')\sum_n  
\psi_{\theta,n}(\phi_f)\psi_{\theta, n}^*(\phi_i) \ee^{-\ii E^{(\theta)}_n (t_f-t_i)/\hbar}
=\delta(\theta-\theta')
\left.\langle \phi_f, t_f| \phi_i, t_i\rangle\right|_{\theta}\ . 
\end{equation}
We can now use the identity
\begin{equation}
\sum_{Q=-\infty}^{\infty} \ee^{-\ii Q(\theta-\theta')}=2\pi\delta(\theta-\theta')
\end{equation}
to select the value of $\theta$ we are interested in: using
Eq.~\eqref{eq:thetabc} and Eq.~\eqref{eq:pathint} we have 
\begin{equation}\label{eq:qm_fixtheta}
\begin{aligned}
&\left.\langle \phi_f, t_f| \phi_i, t_i\rangle\right|_{\theta}  = \int_0^{2\pi}\dd\theta' \delta(\theta-\theta')\sum_n  
\psi_{\theta',n}(\phi_f)\psi_{\theta', n}^*(\phi_i) \ee^{-\ii E^{(\theta')}_n (t_f-t_i)/\hbar}
= \frac{1}{2\pi}\int_0^{2\pi}\dd\theta' 
\sum_{Q=-\infty}^{\infty} \ee^{-\ii Q(\theta-\theta')}\sum_n \psi_{\theta',n}(\phi_f)\psi_{\theta', n}^*(\phi_i) \ee^{-\ii E^{(\theta')}_n (t_f-t_i)/\hbar}= \\ 
&\qquad=\frac{1}{2\pi} \sum_{Q=-\infty}^{\infty} 
\ee^{-\ii Q\theta}\int_0^{2\pi}\dd\theta' 
\sum_n \psi_{\theta',n}(\phi_f+2\pi Q)\psi_{\theta', n}^*(\phi_i) \ee^{-\ii E^{(\theta')}_n (t_f-t_i)/\hbar}= 
\frac{1}{2\pi}\sum_{Q=-\infty}^{\infty} \ee^{-\ii Q\theta} \langle \phi_f+2\pi Q, t_f |\phi_i, t_i\rangle =\\
&\qquad= \frac{1}{2\pi}\sum_{Q=-\infty}^{\infty} \ee^{-\ii Q\theta}
\int_{\phi(t_i)=\phi_i}^{\phi(t_f)=\phi_f+2\pi Q} \mathcal{D}\phi(t)\, \exp\left(\,\frac{\ii}{\hbar}\int_{t_i}^{\,t_f}L(\phi,\dot{\phi})\dd t\right)
\propto\int_{\phi(t_i)\,=\,\phi_i}^{\phi(t_f)\,=\,\phi_f} \mathcal{D}\phi(t) \,\exp\left(\,\frac{\ii}{\hbar}\int_{t_i}^{\,t_f}
\left(L(\phi,\dot{\phi})-\hbar\frac{\theta}{2\pi}\dot{\phi}\right)
\dd t\right)\ ,
\end{aligned}
\end{equation}
where in the last step $\phi_i$ and $\phi_f$ are defined modulo $2\pi$, the
path-integral is performed by summing on all path topologies and we used
$\exp(-\ii\frac{\theta}{2\pi}\int_{t_i}^{t_f}\dot{\phi}\dd t)=\ee^{-\ii
Q\theta}\ee^{-\ii\frac{\theta}{2\pi}(\phi_f-\phi_i)}$. We thus obtain the
Lagrangian
\begin{equation}\label{eq:qmLtheta}
L_{\theta}=\frac{1}{2}mR^2\dot{\phi}^2-V(\phi)-\hbar\frac{\theta}{2\pi}\dot{\phi}\ .
\end{equation}

The last term of the Lagrangian $L_{\theta}$ has the typical form of a $\theta$
term: it is a term of quantum origin (as signaled by the explicit presence of
$\hbar$ in the Lagrangian) which is a total derivative with respect to time,
hence it has no effect on the classical equation of motion of the system, still it
has non trivial effects in the quantum theory. To verify the last statement,
i.~e. the effect of the $\theta$-term in the quantum theory, there are (at
least) two ways. The first, more direct, one consists in explicitly computing
the energy spectrum of the model in the free case $V=0$: it is indeed trivial
to verify that in this case the energy spectrum explicitly depends on $\theta$,
and it is given by
\begin{equation}\label{eq:thetadepfree}
E_n^{(\theta)}=\frac{\hbar^2}{2mR^2}\left(n+\frac{\theta}{2\pi}\right)^2\ , \quad n\in\mathbb{Z}\ .
\end{equation}
The second method uses the path-integral formulation: the $\theta$-term
contributes to the propagator from the state $|\phi_i\rangle$ at time $t_i$ to
the state $|\phi_f\rangle$ at time $t_f$ with  
\begin{equation}
\exp\left(-\frac{\ii\theta}{2\pi}\int_{t_i}^{\,t_f}\dot{\phi}\dd t\right) =
\exp\left(-\frac{\ii\theta}{2\pi}[\phi_f(t_f)-\phi_i(t_i)]\right)\ .
\end{equation} 
The effect of the $\theta$-term is thus of topological origin, as can be seen by
noting that, for fixed $\phi_i(t_i)$ and $\phi_f(t_f)$, the contributions of two
different paths have a relative weight $\exp(-\ii\theta\Delta Q)$, where $\Delta
Q$ is the difference between the winding numbers along the circumference of the
two paths. Note that the winding number of two open (i.e., with
$\phi_f\neq\phi_i$) paths is not well defined, but their difference is. For the
particular case of the quantum particle moving on a circumference we can also
note that $\theta$ enters the Lagrangian $L_{\theta}$ in the same way as a
constant vector potential directed along the $\phi$ direction:
\begin{equation}
L_{A_{\phi}}=\frac{1}{2}mR^2\dot{\phi}^2-V(\phi)-\frac{e}{c}R\dot{\phi}A_{\phi}\ , 
\end{equation}
and by introducing the magnetic flux $\Phi_B=2\pi RA_{\phi}$ we can write
\begin{equation}
\frac{\theta}{2\pi}=\frac{\Phi_B}{\Phi_0}\ , 
\end{equation}
where we defined the elementary magnetic flux $\Phi_0=\frac{2\pi\hbar c}{e}$.
For this simple model $\theta$-dependence is thus just a manifestation of the
Ahronov-Bohm effect, see, e.~g., \cite{BallentineBook, KonishiPaffutiBook}.

We are now ready to discuss the case of four-dimensional gauge theories,
considering for the sake of the simplicity the case of the gauge group SU($N$)
and using natural units. The Lagrangian
density of the pure glue theory is 
\begin{equation}\label{eq:FF}
\mathcal{L}=-\frac{1}{4}F_{\mu\nu}^aF^{a\,\mu\nu} 
\end{equation} 
where
$F^a_{\mu\nu}=\partial_{\mu}A_{\nu}^a-\partial_{\nu}A_{\mu}^a+gf^{abc}A_{\mu}^bA_{\nu}^c$
is the $a$-th component of field strength tensor, $a=1,\ldots, N^2-1$ denotes
the color component, $f^{abc}$ are the SU($N$) structure constants, and $g$ is
the gauge coupling. The possible presence of matter fields does not
significantly affect the following discussion, except for some details that will be
signaled in due time.  The first step required to start the canonical
quantization programme is to write the Hamiltonian density corresponding to the
Lagrangian density in Eq.~\eqref{eq:FF} (something which is in fact required
also in the path integral construction), and we immediately face a well known
problem: the variable $\dot{A}^0$ does not enter Eq.~\eqref{eq:FF}, so we can
not define its conjugate momentum $P_0$.  The theory of constrained Hamiltonian
systems has been developed, starting from Dirac \cite{DiracBook}, to deal with
singular Lagrangians, i.~e. those cases in which the Legendre transformation
between momenta and velocities is not well defined. A simple account of the
basic ideas can be found in \cite{SmilgaBook}, while more details are
presented, e.~g., in \cite{Weinberg1Book, HansonReggeTeitelboimBook,
HenneauxTeitelboimBook}, and the outcome of this analysis is that it is
legitimate to consider the spatial component $A_j^a$, $j=1,2,3$ and their
conjugate momenta $P_j^a$ as dynamical variables subject to the constraint
given by the equation of motion obtained by varying $\mathcal{L}$ with respect
to $A^0$. In particular it is simple to verify that $P_j^a=F_{0j}^a$ and that
the constraint can be written in the form
\begin{equation}\label{eq:G}
G^a\equiv\partial_j P_j^a+gf^{abc}A_j^bP_j^c=0\ .
\end{equation}
Note that, since the structure constants are antisymmetric in all indices, there
is no problem related to the operator ordering in the second term of the
previous expression. 

To understand the physical meaning of the constraint in Eq.~\eqref{eq:G} we have
to remember how gauge transformations acts on the gauge field: if we introduce
for the sake of the simplicity the matrix notation $A_{\mu}=A_{\mu}^aT^a$ (where
$T^a$ are the Hermiatian generators of the fundamental SU($N$) representation,
normalized by $\mathrm{Tr}(T^aT^b)=\frac{1}{2}\delta^{ab}$) a finite gauge
transformation $\Omega(x)$ acts on the gauge field as follows 
\begin{equation}
A_{\mu}\to A_{\mu}^{\Omega}=\Omega(x)A_{\mu}\Omega^{\dag}(x)-\frac{\ii}{g}[\partial_{\mu}\Omega(x)]\Omega^{\dag}(x)\ , 
\end{equation} 
and in particular, by writing $\Omega(x)=\ee^{\ii
gT^a\chi^a(x)}\approx 1+\ii gT^a\chi^a(x)$, the change of the spatial components
of the gauge field under an infinitesimal gauge transformation is
\begin{equation} 
\delta A_j^a=\partial_i \chi^a-gf^{abc}\chi^b A_j^c\ .
\end{equation} 
All gauge transformations that will be used in the following will
be assumed to be local, i.~e. $\Omega(\mathbf{x})\to 1$ sufficiently fast as
$|\mathbf{x}|\to\infty$ (we denote by $\mathbf{x}$ the spatial components of
$x$, at fixed time); this requirement is needed in order to disentangle
the local transformations from the global ones, that are associated to the conserved
charges. Infinitesimal gauge transformations, in particular, have to satisfy
$\chi^a(\mathbf{x})\to 0$ as $|\textbf{x}|\to \infty$. If we now denote by
$\Psi[\mathbf{A}]$ the wavefunction of the system (with $\textbf{A}$ denoting
collectively the spatial components $A_j(\mathbf{x})$ for a fixed time), the
variation of $\Psi[\mathbf{A}]$ due to an infinitesimal gauge transformation is
(integrating by part and using $P^{a\, j}(x)=-\ii\frac{\delta}{\delta
A_j^{a}(x)}$) 
\begin{equation}
\Psi[\mathbf{A}+\delta \mathbf{A}]-\Psi[\mathbf{A}] \simeq \int
\frac{\delta\Psi[\mathbf{A}]}{\delta A_j^a}\delta A_j^a\dd^3\mathbf{x}= -\int
\left(\chi^a\partial_j\frac{\delta}{\delta A_j^a}+gf^{abc}\chi^a
A_j^b\frac{\delta}{\delta A_j^c}\right) \Psi[\mathbf{A}]\dd^3\mathbf{x} \propto
\int \chi^a(\mathbf{x}) G^a\Psi[\mathbf{A}]\dd^3\mathbf{x}\ .  
\end{equation}
Since $\chi^a(\mathbf{x})$ approaches zero for large $\textbf{x}$ we
could integrate by part in the second step of the previous equation, and the
final integral is well defined. We thus see that the constraint in
Eq.~\eqref{eq:G} implements gauge invariance under infinitesimal local gauge
transformations. If
matter is coupled to the gauge field, a further term appears in
Eq.~\eqref{eq:G}, related to the variation with respect to $A^0$ of the matter
part of the Lagrangian density.  The resulting constraint once again implements
gauge invariance, this time under infinitesimal local gauge transformations of
both gauge and matter fields.

Not all finite local gauge transformations can however be continuously
connected to the identity by infinitesimal transformations. Since
$\Omega(\mathbf{x})$ approaches the identity as $|\textbf{x}|\to \infty$,
$\Omega(\mathbf{x})$ effectively defines a map from the 3-sphere $S^3$ (which
is the one-point-compactification of $\mathbb{R}^3$, obtained by adding the
point at infinity) to the gauge group, and for SU($N$) the third homotopy group
is nontrivial: $\pi_3(\mathrm{SU}(N))=\mathbb{Z}$. This can be made more
concrete by introducing the quantity
\begin{equation}
N(\Omega)=\frac{1}{24\pi^2}{\epsilon}^{ijk}\int
\mathrm{Tr}\left[\Omega^{-1}(\partial_i \Omega) \Omega^{-1}(\partial_j \Omega) \Omega^{-1}(\partial_k \Omega)\right]\dd^3\mathbf{x}\ .
\end{equation}	
It is indeed possible to show (see, e.~g., \cite{Weinberg2Book, WeinbergEBook})
that $N(\Omega)$ is the winding number of the mapping $\Omega(\mathbf{x})$ from
$S^3$ to SU($N$). In particular $N(\Omega)$ is invariant under infinitesimal
changes of $\Omega$, and $N(\Omega_1\Omega_2)=N(\Omega_1)+N(\Omega_2)$. It is
also possible to show, by explicit computation, that the winding number of the
transformation 
\begin{equation}
\Omega_1=\exp\left(\ii \frac{x_j\sigma_j}{|\mathbf{x}|}f(|\textbf{x}|)\right)\ ,\
\end{equation}
where $f$ is a monotonic function with $f(0)=-\pi$ and $f(\infty)=0$, is
$N(\Omega_1)=1$, hence we have an explicit example of a gauge transformation
with nonvanishing winding number. Such topologically nontrivial gauge
transformations are often called ``large'' gauge transformations, as opposed to
the ``small'' ones, which are continuously connected to the identity.

The large gauge transformation $\Omega_1$ defined
above is the equivalent of the $2\pi$ rotation for the quantum particle on a
circumference. If we denote by $U$ the unitary operator implementing the gauge
transformation $\Omega_1$ in the Hilbert space of the theory, since the theory
is gauge invariant $U$ commutes with the Hamiltonian, but the constraint in
Eq.~\eqref{eq:G} is not sufficient to fix $U=1$.  The best we can do is to
simultaneously diagonalize $U$ and the Hamiltonian, once again fixing the
eigenvalue $\ee^{\ii\theta}$ of the unitary operator $U$. In terms of the
wavefunction $\Psi(\mathbf{A})$ this amounts to impose
$\Psi(\mathbf{A}^{\Omega_1})=\ee^{\ii\theta}\Psi(\mathbf{A})$, which
implies, given the properties of $N(\Omega)$ and the invariance under
infinitesimal transformations of $\Psi$, the more general
\begin{equation}
\Psi(\mathbf{A}^{\Omega})=\ee^{\ii\theta N(\Omega)}\Psi(\mathbf{A})\ .
\end{equation}

We are now almost ready to adapt the reasoning carried out in
Eq.~\eqref{eq:qm_fixtheta} to the case of four-dimensional gauge theories. The
last ingredient needed is a relation between the winding number $N(\Omega)$ and
four-dimensional integrals.  To this purpose we introduce the (non gauge
invariant) Chern--Simons current
\begin{equation}
j_{\scriptscriptstyle{\rm CS}}^{\mu}=\frac{g^2}{16\pi^2}\epsilon^{\,\mu\alpha\beta\gamma}\mathrm{Tr}\left(F_{\alpha\beta}A_{\gamma}
+\frac{2\ii g}{3}A_{\alpha}A_{\beta}A_{\gamma}\right)\ ,
\end{equation}
its gauge invariant four-divergence (usually called topological charge density)
\begin{equation}\label{eq:q}
q\equiv \partial_{\mu}j_{\scriptscriptstyle{\rm CS}}^{\mu}=\frac{g^2}{32\pi^2}\epsilon^{\mu\nu\rho\sigma}\mathrm{Tr}\left(F_{\mu\nu}F_{\rho\sigma}\right)
=\frac{g^2}{64\pi^2}\epsilon^{\,\mu\nu\alpha\beta}F^a_{\mu\nu}F^a_{\alpha\beta}\ ,
\end{equation}
and its integral on a fixed time surface
\begin{equation}
W(\mathbf{A})=\int j_{\scriptscriptstyle{\rm CS}}^0(t, \textbf{x})\dd^3\textbf{x}\ .
\end{equation}
Using the four-dimensional divergence theorem and assuming the spatial
components of the Chern--Simons current to be trivial at spatial infinity (e.~g.
due to some fixed boundary conditions) we have
\begin{equation}
\int q(x) \, \dd^4x=\int (\partial_0j^0_{CS}+\partial_k j^k_{CS})\dd^4x=
\int_{t_i}^{t_f} \dd t \, \partial_0 W(\mathbf{A})=W(\textbf{A}_f)-W(\textbf{A}_i)\ ,
\end{equation}
moreover under the gauge transformation $\mathbf{A}\to \textbf{A}^{\Omega}$ it
is possible to check by direct computation \cite{Jackiw85} that $W(\mathbf{A})$
transforms as follows
\begin{equation}\label{eq:WOmega}
W(\mathbf{A}^{\Omega})=W(\textbf{A})-N(\Omega)\ .
\end{equation}
We can now repeat almost verbatim Eq.~\eqref{eq:qm_fixtheta}: 
\begin{equation}
\begin{aligned}
\left. \langle \mathbf{A}_f, t_f| \textbf{A}_i, t_i\rangle\right|_{\theta}&=
\int_0^{2\pi}\dd\theta'\frac{1}{2\pi}\sum_{\nu=-\infty}^{\infty}\ee^{-\ii\nu(\theta-\theta')}
\sum_n \Psi_{\theta', n}(\mathbf{A}_f) \Psi_{\theta', n}(\textbf{A}_i)\ee^{-\ii E_n^{(\theta')}(t_f-f_i)}=\\
&=\frac{1}{2\pi}\sum_{\nu=-\infty}^{\infty}\ee^{-\ii\nu\theta} \int_0^{2\pi}\dd \theta'
\sum_n \Psi_{\theta', n}(\mathbf{A}_f^{\Omega_1^{\nu}}) \Psi_{\theta', n}(\textbf{A}_i)\ee^{-\ii E_n^{(\theta')}(t_f-f_i)}
=\sum_{\nu=-\infty}^{\infty}\ee^{-\ii\nu\theta}\langle \mathbf{A}_f^{\Omega_1^{\nu}}, t_f| \textbf{A}_i,t_i\rangle =\\
&=\frac{1}{2\pi}\sum_{\nu=-\infty}^{\infty}\ee^{-\ii\nu\theta}\int_{\mathbf{A}(t_i)=\textbf{A}_i}^{\textbf{A}(t_f)=\textbf{A}_f^{\Omega_1^{\nu}}}
\mathcal{D}A(t,x)\,\exp\left(\ii\int_{t_i}^{t_f}\dd t\int \dd x \mathcal{L}\right)
\propto\int_{\mathbf{A}(t_i)=\textbf{A}_i}^{\textbf{A}(t_f)=\textbf{A}_f}\mathcal{D}A(t,x)\,
\exp\left(\ii\int_{t_i}^{t_f}\dd t\int \dd x \left[\mathcal{L}+\theta q\right]\right)\ ,
\end{aligned}
\end{equation}
where in the last step we used the gauge invariance of $\mathcal{L}$ under time dependent gauge transformations and  
\begin{equation}
\int_{t_i}^{t_f}\dd t\int\dd \mathbf{x}\, q(t,\mathbf{x})=W(\mathbf{A}_f^{\Omega_1^{\nu}})-W(\mathbf{A}_i)=
W(\mathbf{A}_f)-W(\mathbf{A}_i)-\nu\ .
\end{equation}

\subsection{The strong CP problem}

The Lagrangian density of QCD, taking into account also the
matter fields and the $\theta$-term, is thus
\begin{equation}
\mathcal{L}=-\frac{1}{4}F^a_{\mu\nu}F^{a\,\mu\nu}+\sum_{\subf} \bar{\psi}_{\subf}(\ii\gamma_{\mu}D^{\mu}-m_{\subf})\psi_{\subf}
+\theta\frac{g^2}{64\pi^2}\epsilon^{\,\mu\nu\alpha\beta}F^a_{\mu\nu}F^a_{\alpha\beta}\ ,
\end{equation}
where $D_{\mu}=\partial_{\mu}-\ii g A_{\mu}^aT^a$ is the covariant derivative in
the fundamental representation of the gauge group and ``f'' is a flavor index.
It is well known that the first two terms of this Lagrangian density (just like
in Quantum Electrodynamics) are invariant under both parity (P) and charge
conjugation (C), however the $\theta$-term explicitly breaks P and CP
for generic values of the $\theta$ angle, and this is essentially due to the
presence of the Levi-Civita totally antisymmetric tensor. If we introduce the
three dimensional chromoelectric and chromomagnetic fields in analogy with
classical electromagnetism we have indeed that $q\propto
\mathbf{E}^a\cdot\mathbf{B}^a$, and since the chromomagnetic field is even under
parity transformation we see that the $\theta$-term is odd under both P and
CP. Invariance under parity and CP is clearly recovered for $\theta=0$, but
also for $\theta=\pi$: since $\theta$ is defined up to $2\pi$, we have indeed
$\pi\sim-\pi$. Note that at $\theta=\pi$ the discrete symmetries are however
expected to be spontaneously broken, since at $\theta=\pi$ a level crossing
happens, just like in Eq.~\eqref{eq:thetadepfree} (see also the following
discussion on $\theta$-dependence in chiral perturbation theory in
Sec.~\ref{sec:largeN_chiPT} and \cite{Witten:1980sp,Gaiotto:2017yup}).

Parity violations in strong interactions are (if they exist at all) very small,
so it is meaningful to perform an expansion in $\theta$, which parametrizes the
only term in the Lagrangian density which explicitly breaks P and CP
symmetries. Violations of these symmetries in physical observables are thus
expected to be linear in $\theta$ with good accuracy.  The
neutron electric dipole moment $d_{\rm n}$ has been identified as one of the most
sensitive observables that can be used to identify violation of P and CP in
strong interactions, and using dimensional analysis one gets the estimate
\begin{equation}\label{eq:NEDM_dimensional}
|d_{\rm n}| \sim  |\theta| \, e\, \frac{m_{\pi}^2}{m^3_{\rm n}}\ ,
\end{equation}
where $e$ is the elementary electric charge, and $m_{\pi}$ and $m_{\rm n}$
are the pion and nucleon mass, respectively.  Starting from \cite{Baluni:1978rf,
Crewther:1979pi} several attempts have been made to improve this dimensional
estimate using phenomenological models or chiral expansions, and the latest
estimate from chiral perturbation theory ($\chi$PT) yields \cite{Guo:2012vf}
\begin{equation}
d_{\rm n} = (-2.9\pm 0.9) \cdot 10^{-3} \, \theta \, e \, \mathrm{fm}\ .
\end{equation}
The determination of the $\theta$ electric-polarizability of the neutron 
by using lattice QCD methods is a notoriously difficult task, and the values estimated in this way are
in the same ballpark of those obtained by using $\chi$PT, see \cite{Liu:2024kqy} for a recent
review.
A comparison with the experimental measure \cite{Abel:2020pzs}
\begin{equation}
| d_{\rm n} | = (0.0\pm1.1_{\rm stat} \pm 0.2_{\rm syst})\cdot 10^{-13} \, e \, \mathrm{fm}
\end{equation}
gives the upper bound $|\theta| \lesssim 10^{-10}$ for the value of the
$\theta$ parameter. So why not fixing directly $\theta=0$ in
$\mathcal{L}_{\theta}$? If we only considered QCD,
independently of the other sectors of the Standard Model, this would indeed be
possible: it would be enough to impose from the beginning local gauge
invariance, renormalizability, and invariance under parity to obtain
$\mathcal{L}_{\theta}$ with $\theta=0$. 

To discuss why this is not possible when considering the whole Standard Model
we have to introduce one last ingredient: the chiral anomaly
\cite{Adler:1969gk, Bell:1969ts}. If we consider a four-dimensional gauge
theory coupled to a single flavor of quark, the chiral rotation
\begin{equation}\label{eq:chiraltrans}
\psi\to \ee^{\ii\alpha\gamma^5}\psi\ ,\quad \bar{\psi}\to \bar{\psi}\ee^{\ii\alpha\gamma^5}
\end{equation}
can be shown to be equivalent to a complex rotation of the mass 
and to a shift of the $\theta$ angle (see, e.~g., \cite{Weinberg2Book}):
\begin{equation}
m\bar{\psi}\psi \to
\bar{\psi}m'\frac{1+\gamma^5}{2}\psi +\bar{\psi}m'^*\frac{1-\gamma^5}{2}\psi\ ,\quad m'=m \ee^{2\ii\alpha}\ ,\quad
\theta\to \theta+2\alpha\ .
\end{equation}
This relation implies that we can reabsorb $\theta$ in the mass term at the
expense of using the complex mass $m \ee^{-\ii\theta}$. Note that this does not
change the invariance properties of the theory, since a complex mass is not
invariant under P and CP transformation, but gives us the possibility of
studying $\theta$-dependence using chiral perturbation theory if light quarks
are present (and this is the rationale behind Eq.~\eqref{eq:NEDM_dimensional},
since $m_{\pi}^2\propto m$).  An immediate consequence of this fact is that
there is no $\theta$-dependence at all if $m=0$, since in this case we can
shift $\theta$ at will without affecting the mass term of the Lagrangian.  When
both the $\theta$-term and a complex mass are present, we can define the
physical $\theta$ angle as the combination $\theta_{\scriptscriptstyle{\rm phys}}=\theta-\arg m$, which
is invariant under chiral transformations and corresponds to the value of the
$\theta$ angle when the fermion mass is real and positive.  The previous
discussion can be generalized to the case of a generic number of flavors with
mass term
\begin{equation}
\bar{\psi}M\frac{1+\gamma^5}{2}\psi +\bar{\psi}M^{\dag}\frac{1-\gamma^5}{2}\psi\ ,
\end{equation}
where $M$ is a $N_{\subf}\times N_{\subf}$ matrix and the physical $\theta$ angle is in
this case given by
\begin{equation}
\theta_{\scriptscriptstyle{\rm phys}}=\theta-\arg\det M\ .
\end{equation}

The previous statement regarding the upper limit of the $\theta$ angle should
thus more precisely be written as $|\theta_{\scriptscriptstyle{\rm phys}}|\lesssim 10^{-10}$, and the
presence of $\theta_{\scriptscriptstyle{\rm phys}}$ shows why we can not simply fix to zero the
$\theta$ angle in the Standard Model: the mass matrix originates from the
electroweak sector, and since in the electroweak sector P and CP are not
conserved, there is no way to fix $\theta_{\scriptscriptstyle{\rm phys}}=0$ using symmetry
arguments. The fact that $|\theta_{\scriptscriptstyle{\rm phys}}|$ is so small or, equivalently, the
absence of P and CP violations in the strong sector of the Standard Model,
constitutes the so called strong CP problem.  As noted before there is no
$\theta$-dependence at all if at least one quark is massless ($\arg \det M$ is
not defined if $M$ is singular). In the past, a massless up quarks has thus been
suggested as a simple solution of the strong CP problem, however this is today
known to be inconsistent with experimental and lattice QCD results
\cite{FlavourLatticeAveragingGroupFLAG:2021npn, Sanfilippo:2015era,
ParticleDataGroup:2024cfk} (see also \cite{Alexandrou:2020bkd}).

%% file: sec3_res.tex
\section{General aspects of \texorpdfstring{$\theta$}{theta}-dependence}
\label{sec3}

As we argued above, within the general framework of the Standard Model one
cannot exclude \emph{a priori} a $\theta$-term in the QCD Lagrangian.  At the same
time, the dependence of various properties of strong interactions on $\theta$
plays a significant role by itself, which is relevant to many phenomenological
and theoretical issues, like the interpretation of some particular features of
the hadronic spectrum, or the characterization of the various phases which are
found when considering Yang--Mills theories or QCD at finite temperature.
Moreover, one of the proposed solutions to the strong CP problem is the
so-called Peccei--Quinn mechanism~\cite{Peccei:1977hh,Peccei:1977ur}, which
invokes the presence of a new particle, the axion, which has also become a
popular candidate for dark matter: since the axion couples to QCD like a
$\theta$-term, the study of $\theta$-dependence in QCD turns out to be
fundamental in order to predict some properties of this dark matter candidate.

The purpose of this Section is to go more in detail into these interesting
aspects of $\theta$-dependence, and to review the knowledge that we have
achieved so far. It should be clear that, being linked to the existence of
large field fluctuations with non-trivial topology, the problem is
non-perturbative, hence the present knowledge is based either on general
non-perturbative theoretical approaches to Yang--Mills theories, like, e.~g., the
large-$N$ expansion, or on numerical computations based on the lattice
formulation.

The discussion about $\theta$-dependence is best formulated within the
Euclidean path-integral formulation of the theory, which provides the framework
for the lattice QCD numerical approach, and is also the natural setting for the
finite temperature theory. Therefore, in order to fix ideas, we consider the
path-integral formulation of the QCD thermal partition function which, up to an
overall multiplicative normalization constant, reads:
\beq
Z(\theta,T) = \int \mathcal{D} A \, \mathcal{D} \overline{\psi} \, \mathcal{D} \psi \, \ee^{-\mathcal{S}_{\subE}+\ii\theta Q}
= \int \mathcal{D} A \prod_{\mathrm{f}\,=\,1}^{N_{\subf}}\det\left[\slashed{D}[A]+m_{\subf}\right] \, \ee^{-\mathcal{S}_{\subE}^{\supYM} + \ii \theta Q}
\label{eq:Z_thermal}
\eeq
where $\mathcal{S}_{\subE} = \int \mathrm{d}^4x\, \mathcal{L}_{\subE}$
is the Euclidean QCD action,
$Q = \int \mathrm{d}^4x\, q(x)$, and
\begin{equation}
  \mathcal{L}_{\subE} = \frac{1}{4}F^a_{\mu\nu}F^{a}_{\mu\nu}+\sum_ {\subf} \bar{\psi}_{\subf}(\gamma^{\subE}_{\mu}D_{\mu}+m_{\subf}) \psi_{\subf} \ ;
  \ \ \ \ \ \ \ 
  q(x) = \frac{g^2}{32\pi^2} F^a_{\mu\nu} \tilde F^a_{\mu\nu}\ ; \ \ \ \ \ \ \ \tilde F^a_{\mu\nu} =
  \frac{1}{2} \epsilon_{\,\mu\nu\alpha\beta} F^a_{\alpha\beta} \, ,  
\end{equation}
where we have introduced the dual field strength tensor $\tilde F^a_{\mu\nu}$,
while $\mathcal{S}_{\subE}^{\supYM}$ refers to the pure gauge part of
$\mathcal{S}_{\subE}$.  The Euclidean Lagrangian is defined in terms of the
Wick rotated fields, $t \to -\ii \tau$ where $\tau$ is the Euclidean time, and of
the Euclidean gamma matrices ($\gamma_0^{\subE} = \gamma^0 \ , \ \gamma_j^{\subE} = - \ii
\gamma^j$).  Space-time integration is assumed to be compactified in the
Euclidean time direction, with a compactification extension $\beta_T = 1/T$,
where $T$ is the physical temperature, and periodic/antiperiodic boundary
conditions in the time direction for bosonic/fermionic fields, in order to
reproduce thermal conditions.

In Eq.~(\ref{eq:Z_thermal}), an exact partial integration over the fermion
fields has been performed, leading to a path-integral formulation purely in
terms of bosonic (gauge) fields, where the information about fermions is
encoded in the determinants of the fermion kernels appearing after integration.
This is a necessary step for a numerical computation of the partition function,
since in the path-integral formulation fermions are represented by Grassmann
variables, which are not usual $c$-numbers. In the lattice QCD approach, after
proper discretization, the path-integral is sampled by Monte Carlo algorithms:
this is possible for $\theta=0$ thanks to the positivity of
$\prod_{\mathrm{f}\,=\,1}^{N_{\subf}}\det\left[\slashed{D}[A]+m_{\subf}\right]
\, \ee^{-\mathcal{S}_{\subE}^{\supYM}}$, which can then be interpreted as a
probability distribution function over the gauge fields, the positivity of the
determinants being usually guaranteed by the properties of the fermion kernels.
Thermal expectation values can then be computed as average values of proper
field functionals over this distribution; as $\beta_T \to \infty$, the thermal
boundary conditions become irrelevant and one recovers vacuum expectation
values.

It is clear from Eq.~(\ref{eq:Z_thermal}) that, in the Euclidean formulation,
the $\theta$-term gives a purely imaginary contribution to the action, making
it complex.  This is usually known as then {\em sign problem}, since it ruins
the positivity of the Euclidean path-integral measure, thus undermining the
Monte Carlo approach at its very basis. A similar problem appears in other
cases, for instance, the fermion determinants themselves becomes complex after
adding chemical potentials coupled to quark numbers.  In this case, its origin
can be traced back to the fact that the $\theta$-term involves an odd number of
time derivatives (contrary to the standard pure gauge term), hence the
appearance of the extra factor ``$\ii$''. 

That also underlies one of the exact results about
$\theta$-dependence~\cite{Vafa:1983tf,Vafa:1984xg}, the so-called Vafa--Witten
theorem.  Let us consider the $\theta$-dependent part of the free energy
density (vacuum energy density at $T = 0$):
\beq
F(T,\theta) = -\frac{1}{\mathcal{V}}\log\left[\frac{Z(T,\theta)}{Z(T,0)}\right]
\label{eq:Ftheta}
\eeq
where $\mathcal{V}$ is the four-dimensional volume. The extra factor ``$\ii$''
of the Euclidean $\theta$-term implies that the ratio of partion functions can be viewed as the expectation
value of a pure  phase factor, $Z(T,\theta)/Z(T,0) = \braket{\ee^{\ii \theta
Q}}_{\theta = 0}$, which is then bounded by 1 from above, meaning that $F(T,\theta)
\geq 0$ with a minimum in $\theta = 0$\footnote{The same property holds for the
quantum particle moving on a circumference discussed above: in that case, since
a non-zero $\theta$ can be interpreted as the presence of a magnetic flux, the
property implies that the system is diamagnetic.} That rules out the
possibility of spontaneous P breaking at $\theta = 0$~\cite{Vafa:1984xg}, hence
$F(T,\theta)$ is an even function of $\theta$ with period $2 \pi$ (because $Q
\in \mathbb{Z}$) and analytic around $\theta = 0$. 
Therefore, $F(T,\theta)$ can be written as a Taylor series around $\theta = 0$,
containing only even powers.  A common parametrization is the following:
\beq\label{eq:thetadep_free_energy_general}
F(T,\theta) = \frac{1}{2}\chi\theta^2 \left(1+\sum_{n\,=\,1}^{\infty}b_{2n}\theta^{2n}\right) \, .
\eeq
Eq.~(\ref{eq:Ftheta}) and the relation $Z(T,\theta)/Z(T,0) = \braket{\ee^{\ii
\theta Q}}_{\theta = 0}$ permit to rewrite the expansion coefficients in terms
of expectation values computed at $\theta = 0$
\beq\label{eq:chib2n_def}
\chi = \frac{1}{\mathcal{V}}\braket{Q^2} \ ; \ \ \ \ \ 
b_{2n} = \frac{2(-1)^{n}}{(2n+2)!} \frac{\braket{Q^{2n}}_{c}}{\braket{Q^2}} \ ,
\eeq
where
$\braket{\cdot}_c$ stands for a connected expectation value, e.~g.,
$\braket{Q^4}_c = \braket{Q^4} - 3 \braket{Q^2}^2$.

A nice feature of this parametrization of $F(T,\theta)$ is that all information
is encoded in expectation values which are taken at $\theta = 0$, therefore
lattice simulations can be effectively used to compute them, without facing any
sign problem.  The leading term coefficient, $\chi$, is the so-called
topological susceptibility: as discussed in  the following, its value in the
large-$N$ limit is related to the mass of the $\eta'$ meson by the
Witten--Veneziano mechanism, while its behaviour in the chiral limit is
regulated by Chiral Perturbation Theory ($\chi$PT).  The other coefficients
parameterize higher-order terms in $\theta$ and are proportional to
higher-order cumulants of the topological charge path-integral distribution at
$\theta=0$: they reveal important to distinguish different functional
dependencies of $F(T,\theta)$ on $\theta$ which characterize the possible phases
of Yang--Mills theories.  Note that, given that the free energy density is an
intensive quantity, both $\chi$ and the $b_{2n}$ coefficients have a
well-defined and finite thermodynamic limit. 
\\

As we have discussed above, the problem why $\theta$ is so small does not find
a solution within the Standard Model, given that the possibility of a massless
up quark is ruled out. An attractive solution is provided by the so-called
Peccei--Quinn mechanism~\cite{Peccei:1977hh,Peccei:1977ur, Wilczek:1977pj,
Weinberg:1977ma}, which invokes an extension of the Standard Model with an
extra axial U(1) global symmetry, the U(1)$_{\subPQ}$ Peccei--Quinn symmetry.
Such symmetry is supposed to be spontaneously broken, implying the possible
emergence of a massless Nambu--Goldstone boson, the axion, but also anomalously
broken at the same time, leading to a coupling of the axion field $a(x)$ to the
topological charge density operator $q(x)$.

One can built various theories implementing such features, see e.~g.,
\cite{Kim:1979if, Shifman:1979if, Dine:1981rt, Zhitnitsky:1980tq} for early
proposals and \cite{DiLuzio:2020wdo} for a recent review, however, at the
level of a low energy effective theory, they all correspond to the addition to
the QCD Lagrangian of a massless pseudoscalar with derivative couplings plus a modification of the
$\theta$-term as follows:
\beq
\theta\, q(x) \ \ \to \ \ \left(\frac{a}{f_a} + \theta\right) q(x) 
\label{eq:axion_coupling}
\eeq
where $f_a$ is the axion decay constant, which is supposed to be very large
compared to the Standard Model scales. The key point of the mechanism is that,
in this way, we have an effective $\theta_{\subeff} = \theta + a/f_a$, which
now participates to dynamics.
In particular, the shift invariance $a(x) \to a(x) + \alpha$ that would be
present if U(1)$_{\subPQ}$ were exact is explicitly broken by the anomalous
coupling to $q(x)$, 
so that the axion field, hence
$\theta_{\subeff}$, will select the value which minimizes the vacuum energy
density. By virtue of the Vafa--Witten theorem discussed above, such value is
indeed $\theta_{\subeff} = 0$, which thus solves the CP-problem.

The axion serves multiple scopes: apart from solving the CP-problem, it is also
an interesting dark matter candidate \cite{Preskill:1982cy, Abbott:1982af,
Dine:1982ah}. The interesting point is that, because of its coupling to QCD,
$\theta$-dependence fixes much of its physical properties. For instance, the
second order expansion of the vacuum energy density around the minimum at
$\theta_{\subeff} = 0$ can be interpreted as an effective axion mass term,
leading to a relation between the axion mass $m_a$ and the topological
susceptibility in QCD, $m^2_a = \chi / f_a^2$, while the higher order
coefficients $b_{2n}$ can be related to the axion self couplings.
\\

In the following, we will first review the available predictions and results on
$\theta$-dependence regarding the low temperature phase of SU($N$) gauge
theories, with or without fermions, which is characterized by relevant
phenomena such as color confinement and chiral symmetry breaking, then exposing
predictions and results concerning the high temperature phase, showing how
$\theta$-dependence changes drastically at the transition between the two.

Before doing that, we will briefly review one of the simplest model which can
be used to predict the form of the $\theta$-dependence, known as the Dilute
Instanton Gas Approximation (DIGA) model.  Non-Abelian gauge theories are
characterized by the existence of classical solutions to the equations of
motion, which are either selfdual or anti-selfdual, i.~e., $\tilde F_{\mu\nu} =
\pm F_{\mu\nu}$, with finite action and non-zero topological charge $Q$. The
most elementary of such solutions, with $Q = \pm 1$, are known as instantons
and anti-instantons and carry an action $8 \pi / g^2$, where $g$ is the gauge
coupling: their field strength tensor is non-zero only over a limited portion
of space and time (hence the name), each solution being characterized by a
typical extension radius $\rho$. The reader will find an extensive treatment
about instantons elsewhere in this Volume, as well as in various classical
papers, reviews and textbooks (see, e.~g., Refs.~\cite{Callan:1977gz, Gross:1980br,
Schafer:1996wv, ColemanBook}).

The DIGA model represents a typical semiclassical approach, which assumes that
the dominant contribution to the functional integral can be found by
integrating fluctuations around single (anti)instantons. As we will discuss
later on, such approach is expected to be reliable only at high $T$, while it
fails at low $T$.  There is one good reason to discuss it at this stage: it
clarifies a key point, i.~e., that for each phase of Yang--Mills theories, the
functional dependence of $F(T,\theta)$ is tightly related to the distinctive
topological degrees of freedom which are relevant in that phase.

\subsection{A semiclassical model: the Dilute Instanton Gas Approximation}
\label{sec:diga}

As anticipated above, the DIGA model builds around the hypothesis that the
gauge field configurations giving the most relevant contribution to the
functional integral $Z(T,\theta)$ are those around (i.~e., differing by small
fluctuations) the classical configurations obtained by superposing certain
numbers of instantons (I)  and anti-instanton (A) solutions.  The numbers $n_{\subI} /
n_{\subA}$ of I~/~A solutions for each configuration is arbitrary, however they are
assumed to be dilute enough, i.~e., with typical distances from each other much
greater than their sizes, so that the total Euclidean action can be written as a sum of independent terms, 
and the functional integration of fluctuations around each I~/~A is independent of all the others.
Under such assumptions, and using the fact that the topological charge of each
configuration is $Q = n_{\subI} - n_{\subA}$, the thermal partition function can be written as 
\beq
Z(T,\theta) = \sum_{n_{\subI}\,=\,0}^{\infty} \sum_{n_{\subA}\,=\,0}^{\infty}
\frac{1}{n_{\subI}!n_{\subA}!}Z_1^{n_{\subI}} Z_1^{n_{\subA}}\ee^{\ii\theta(n_{\subI}-n_{\subA})} = \left(\sum_{n_{\subI}\,=\,0}^{\infty} \frac{Z_1^{n_{\subI}}}{n_{\subI}!}\ee^{\ii \theta n_{\subI}}\right) \left(\sum_{n_{\subA}\,=\,0}^{\infty} \frac{Z_1^{n_{\subA}}}{n_{\subA}!}\ee^{-\ii \theta n_{\subA}}\right)
= \exp\left\{\ee^{\ii\theta}Z_1\right\} \exp\left\{\ee^{-\ii\theta}Z_1\right\}
= \exp\left\{2\cos(\theta)\, Z_1\right\}.
\eeq
where $Z_1$ is the single instanton functional integral, obtained by integrating fluctuations
around the single I or A solution, which is actually a function of the
temperature, $Z_1(T)$. The previous computation is in fact identical to 
the one needed to write the classical grand-canonical partition function of two different
species of identical and non-interacting particles, I and A, with $Z_1$ the
one-particle partition function; at $\theta = 0$ one has a Poissonian
distribution for both $n_{\subI}$ and $n_{\subA}$, which is a manifestation of diluteness,
like for a perfect gas.

Using the definition of the free energy in Eq.~(\ref{eq:Ftheta}), as well as
its Taylor expansion representation in
Eq.~(\ref{eq:thetadep_free_energy_general}), one easily finds:
\beq\label{eq:free_energy_theta_DIGA}
F(T,\theta)=-\frac{1}{\mathcal{V}}\log\left[\frac{Z(T,\theta)}{Z(T,0)}\right] = \frac{2Z_1(T)}{\mathcal{V}} \left[1-\cos(\theta)\right] = \chi(T)\left[1-\cos(\theta)\right]
\, .\eeq
It is interesting to notice that the sole hypothesis of diluteness fixes
$F(T,\theta)$, up to an overall pre-factor corresponding to the topological
susceptibility $\chi(T) = {2Z_1(T)}/{\mathcal{V}}$. In particular, one has a
well defined and $T$-independent prediction for the $b_{2n}$ coefficients:
\beq\label{eq:b2n_DIGA}
b_{2n} = \frac{2(-1)^n}{(2n+2)!} \, ,
\eeq
which means, for instance, $b_2 = - 1/12$.

Let us now outline the main facts regarding the computation of $Z_1$, which can
be carried out using perturbation theory on an instanton background
(see, e.~g., Refs.~\cite{Gross:1980br,Pisarski:1980md,Boccaletti:2020mxu} for
more details).  The result contains a contribution proportional to the
exponential of the single instanton action, $\exp( - 8 \pi^2 / g^2(\rho))$,
where $g(\rho)$ is the running coupling computed at the scale of the instanton
size $\rho$, so that the contribution increases/decreases, by asymptotic
freedom, as $\rho$ increases/decreases; such contribution, however, must be
integrated over the instanton size distribution which, for $N$ colors and $N_{\subf}$
flavours turns out to be proportional to $\rho^{\frac{11}{3} N - \frac{2}{3} N_{\subf} - 5}$,
which for $N=3$ diverges for large $\rho$ if $N_{\subf} < 9$. This fact signals the failure of
the DIGA model: while it assumes diluteness, it is dominated by instantons of
arbitrarily large size, which completely spoil diluteness.

However, the outcome changes if large size instantons are suppressed by other
means: this is exactly what happens at finite temperature $T$, which acts as an
infrared cutoff by limiting $\rho \lesssim 1/T$, i.~e., by the extension of the
compactified Euclidean temporal direction. Therefore one expects that, at least
for large enough temperatures (how large, is not known \emph{a priori}) the model
should work. Moreover, for asymptotically large temperatures, one has the
following predictions for the topological susceptibility (assuming $N_{\subf}$ light
flavours of equal mass $m$):
\beq
\chi(T) \propto m^{N_{\subf}} T^{4 - \frac{11}{3} N  - \frac{1}{3}N_{\subf}} \, .
\label{eq:asym_chi_DIGA}
\eeq
Let us stress that the functional dependence for $F(T,\theta)$ predicted by
Eq.~(\ref{eq:free_energy_theta_DIGA}), which is just based on diluteness, could
start to be valid at temperatures lower than those at which the asymptotic
behaviour of Eq.~(\ref{eq:asym_chi_DIGA}), which is the result of a 1-loop
computation, sets in.

\subsection{Low temperature: large-\texorpdfstring{$N$}{N} and chiral scaling}
\label{sec:largeN_chiPT}

The topological properties of QCD in the low temperature phase are strictly
entangled with hadron phenomenology, in particular with features of the hadron
spectrum linked to the chiral anomaly. The most famous example is the role
played by topology in explaining the mass of the $\eta^\prime$
meson~\cite{Witten:1978bc, Witten:1979vv, Veneziano:1979ec, Kawarabayashi:1980dp, Witten:1980sp, DiVecchia:1980yfw}.
The U(1)$_{\subA}$ symmetry that is present in chiral limit of the classical
QCD Lagrangian can not be a quantum symmetry realized \emph{\`a la} Wigner,
since no parity doublet is found in the hadron spectrum. Moreover, as first
pointed out by S.~Weinberg~\cite{Weinberg:1975ui}, the mass of the singlet
pseudoscalar meson $\eta'$ should be smaller than $\sqrt{3}\, m_\pi \simeq
240$~MeV if the U(1)$_{\subA}$ symmetry were spontaneously broken, a bound
which is largely violated by the actual mass, $m_{\eta'} \simeq 958$~MeV.  This is
the so-called U(1)$_{\subA}$ puzzle.  While it became soon clear that the
bound violation must be related to the explicit breaking of U(1)$_{\subA}$ by
the chiral anomaly~\cite{tHooft:1976rip}, the exact mechanism leading to the
phenomenological value of $m_{\eta'}$ is not obvious.

A useful approach to clarify the origin of the $\eta'$ mass is to consider the
limit of QCD known as the 't~Hooft large-$N$ limit~\cite{tHooft:1973alw}, in which we assume
$N \to \infty$ at fixed $N_\subf$, keeping the 't Hooft coupling, $\lambda =
g^2 N$, fixed; the physical interpretation of this condition is that, in this
way, the dynamically-generated hadronic scale $\Lambda_\subQCD$ is kept fixed,
since the QCD $\beta$-function is just a function of $\lambda$ for $N_\subf/N =
0$. The physical mechanism underlying the U(1)$_{\subA}$ puzzle is
identified by noting a peculiarity of the expansion of physical observables in
powers of $1/N$~\cite{tHooft:1973alw}: quark contributions are suppressed by a
factor $1/N$ with respect to gluonic ones, so that in this expansion only
gluons contribute to the zeroth order (i.~e., in the strict large-$N$ limit),
while quarks contributions start from the $O(1/N)$ term.  However, this is at
odds with the fact that $\theta$-dependence should disappear in the presence of
massless quark, i.~e., in the chiral limit, because of the chiral anomaly: this
fact should be valid independently of $N$, however the anomaly is a quark
contribution which should be suppressed at large $N$. 

In order to solve this problem we can hypothesize that, in the
large-$N$ limit, there is a flavor-singlet meson with axial quantum numbers,
the $\eta^\prime$ meson, whose mass squared vanishes at large-$N$ as $1/N$, so
that its propagator at zero momentum is of $O(N)$: that makes $\eta^\prime$'s
contribution leading order in the $1/N$ expansion, thus able to cancel the
gluons' one . In particular, applying this argument to the topological charge
density correlation function, one obtains $m^2_{\eta^{\prime}} = 2 N_{\subf} \,
{\chi^\infty_{_\subYM}} / {F_\pi^2}$ in the chiral limit~\cite{Witten:1979vv}, or the
celebrated Witten--Veneziano equation when the finite quark masses are taken
into account~\cite{Veneziano:1979ec, Witten:1980sp, DiVecchia:1980yfw}:
\beq
\chi_{_\subYM}^\infty = \frac{F_\pi^2}{2 N_\subf} \left( m_{\eta^\prime}^2 + m_{\eta}^2 - 2 m_K^2 \right) \, .
\label{eq:wv}
\eeq
Here, $\chi_{_\subYM}^\infty$ is the topological susceptibility computed in the
pure gauge theory for $N \to \infty$: the right-hand side of Eq.~(\ref{eq:wv})
is of order zero in the $1/N$ expansion, so the Witten--Veneziano mechanism
requires a finite non-zero limit for the pure gauge susceptibility in the $N
\to \infty$ limit which, taking into account phenomenological values of the
meson masses and of the pion decay constant $F_\pi$, turns to be
$\chi_{_\subYM}^\infty \simeq \left(180~\mathrm{MeV}\right)^4$.
\\

A non-zero $\chi_{_\subYM}$ in the $N \to \infty$ limit is at odds with the
DIGA model, which predicts no $\theta$-dependence at all, hence $\chi = 0$, in
the same limit: this is clear from the fact the instanton action $8 \pi^2 / g^2
= 8 N \pi^2 / \lambda$ diverges as $N \to \infty$, so that the 1-instanton
weight vanishes, and is also visible from the asymptotic behavior of $\chi$ in
Eq.~(\ref{eq:asym_chi_DIGA}).
However, we already know that the DIGA model has few hopes to be consistent
but in the high temperature regime, hence we do not expect it to match
low-energy phenomenological predictions.  The point is then whether one can
model and predict the low-temperature $\theta$-dependence in the large-$N$
limit by other means: this can indeed be done on quite general
grounds~\cite{Witten:1980sp,Witten:1998uka}.

On one hand, one expects the free energy density to be proportional to the
number of degrees of freedom, which is $\sim N^2$ in the large-$N$ 't Hooft
limit, since we have $N^2 - 1$ gluons for SU($N$) and fermions become
irrelavant. On the other hand, one expects the relevant variable to be
$\bar{\theta}=\theta/N$ instead of $\theta$, as can be argued observing that,
rescaling the fields by the coupling constant $A^a_{\mu}\to \frac{1}{g}A^a_{\mu}$ to expose the 't Hooft coupling,
the $\theta$-dependent Yang--Mills action can be expressed at large-$N$ as:
\beq
\mathcal{S}^{\supYM}_{\subE}(\theta) = \frac{N}{\lambda} \left[\frac{1}{4} \int \dd^4 x \, F^a_{\mu\nu}(x)F^a_{\mu\nu}(x) -\ii \lambda\frac{\theta}{N} Q\right] \, .
\eeq
Putting these things together, one obtains the following prediction for the
scaling of the free energy density at large-$N$:
\beq
F (T,\theta,N) \underset{N\,\to\,\infty}{\sim} N^2 \bar F \left(T, \frac{\theta}{N}\right) \, .
\label{eq:FlargeNscaling}
\eeq
According to the general parameterization in
Eq.~\eqref{eq:thetadep_free_energy_general}, the scaling function $\bar F$ can
be written as
\beq
\bar F (T, \bar \theta) 
\underset{N\,\to\,\infty}{\sim} \frac{1}{2}\chi(T) \bar{\theta}^2\left(1+\sum_{n\,=\,1}^{\infty} b_{2n} (T) N^{2n}\bar{\theta}^{2n}\right) \, .
\eeq
It is useful to stress that this large-$N$ prediction is not necessarily at odds with the
DIGA model: indeed, they are in perfect agreement if $\chi = 0$, meaning no
$\theta$-dependence at all. However, if $\chi\sim\mathcal{O}(N^0)$ as required
by the Witten--Veneziano mechanism, the large-$N$ argument leads to a
non-trivial prediction, which differs from DIGA substantially. In this case,
indeed, by matching the large-$N$ limit for the scaling function, hence
finite coefficients for the expansion in terms of $\bar \theta$, one can infer
the following scaling behavior for the $b_{2n}$ coefficients (for a large-$N$ analysis of QCD $\theta$-dependence using effective chiral Lagrangians see~\cite{Vonk:2019kwv,GomezNicola:2019myi}):
\beq
b_{2n} \underset{N\,\to\,\infty}{\sim} \frac{1}{N^{2n}} \, ,
\label{eq:largeN_b2N_scaling}
\eeq
so that, in the large-$N$ limit only the leading $\sim\mathcal{O}(\theta^2)$
dependence survives.  In other QCD-like theories characterized by a non-trivial
$\theta$-dependence, like the two-dimensional $\CP^{N-1}$ models, similar
large-$N$ arguments lead to more quantitative predictions for both $\chi$ and
the $b_{2n}$ coefficients~\cite{DAdda:1978vbw, Luscher:1978rn, Witten:1978bc,
Campostrini:1991kv, Campostrini:1991tw, DelDebbio:2006yuf, Rossi:2016uce,
Bonati:2016tvi, Sugeno:2025exv}, essentially because in vector (instead
of matrix) models large-$N$ computations can be explicitly carried out.

However, the semi-quantitative prediction in Eq.~(\ref{eq:largeN_b2N_scaling})
has already quite interesting implications.  In particular, in order for the
analytic dependence on the scaling variable $\theta/N$ to co-exist with the
expected $2\pi$-periodicity in $\theta$, the vacuum energy must have
non-analytic points where level crossing of different branches occur. Thus,
previous arguments suggests the following shape for $F(T,\theta)$ in the
large-$N$ limit:
\beq
F(T,\theta) = \frac{1}{2}\chi(T) \, N^2 \min_{k \, \in \, \mathbb{Z}} \left(\frac{\theta + 2\pi k}{N}\right)^2,
\eeq
which has cusps in $\theta=\pi + 2\pi k$. This non-analyticity suggests that
the CP symmetry is spontaneously broken at $\theta=\pi$ in the large-$N$ limit,
as far as $\chi(T) \neq 0$.
\\

It is interesting to notice that one can try to interpret the different kinds of
$\theta$-dependence which are predicted by the DIGA model on one side, and by
large-$N$ arguments at low temperature on the other side, in terms of the
different degrees of freedom which are thought to be relevant in the two cases.
In the DIGA model, the smooth analytic dependence on $\theta$ emerges by
assuming that the relevant objects are instantons and anti-instantons, which
carry an integer topological charge. In the non-interacting hypothesis, in
particular, one obtains $F(T,\theta) \propto (1 - \cos \theta)$; one could
actually modify such hypothesis by adding interactions, like when moving from a
perfect to a real gas in the virial expansion approach, but still obtain a
smooth function of $\theta$ if the interaction is weak enough.

It is then tempting to interpret the smooth dependence on $\theta/N$ (instead
of $\theta$) which is found in the large-$N$ approach as evidence that, in this
case, the relevant objects carry a fractional topological charge, in particular
in units of $1/N$. Indeed such objects, sometimes called instanton-quarks, and
their possible role in the confined phase have been conjectured since
long~\cite{Fateev:1979dc, Berg:1979uq, Belavin:1979fb,
Gonzalez-Arroyo:1995ynx,Gonzalez-Arroyo:1995isl,Kraan:1998sn,
Zhitnitsky:2006sr, Diakonov:2007nv, Unsal:2008ch, Parnachev:2008fy,
Zhitnitsky:2008ha, Gorsky:2009me, Nair:2022yqi} (see also the recent
review~\cite{Gonzalez-Arroyo:2023kqv} and references therein). In this
perspective, a non-perturbative determination of $\theta$-dependence can
provide information about their dynamics: for instance, if the large-$N$ QCD
vacuum could be described in terms of a non-interacting ensemble of such
objects (like for instantons in the DIGA model), then one would expect
$F(T,\theta) \propto [1 - \cos (\theta/N)]$, hence $b_2 = - 1 / (12 N^2)$. In
the following we will compare this possibility with present lattice results
about $b_2$ in the large-$N$ limit.  \\

Another important aspect of the solution to the U(1)$_{\subA}$ provided by the
Witten--Veneziano mechanism, is that the chiral and the large-$N$ limit do not
commute. Indeed:
\beq
\lim_{m\to 0} \lim_{N\to \infty} \, \chi &=& \chi_{_\subYM}^{\infty} \neq 0 \, ,\\
\lim_{N\to \infty} \lim_{m\to 0} \, \chi &=& 0 \, .
\eeq
So, let us now discuss the predictions that can be obtained in the opposite
limit, $m \to 0$ at fixed $N$, which represents important features of QCD at
the physical point, given the small masses of the light quarks.

As we have discussed above, the chiral anomaly permits to transfer the $\theta$
parameter, by a chiral rotation, from the pure gauge sector to the quark mass
matrix. Then, in the low-temperature phase of QCD characterized by chiral
symmetry breaking, i.~e., when $T\lesssim T_{\rm pc}$, with $T_{\rm pc}\simeq 155$ MeV the
chiral crossover temperature, $\theta$-dependence can be reliably investigated
using chiral perturbation theory ($\chi$PT)~\cite{DiVecchia:1980yfw}.  In particular, for $N_{\subf}=2$
light flavors and at the leading order (LO) in the momentum and quark mass
expansion, the $\theta$-dependent chiral Lagrangian reads:

\beq
\mathcal{L}_{\eff}(\theta)[U] = \frac{F_\pi^2}{4}\Tr\left[\partial_\mu U^{\dag} \partial^\mu U\right] + \frac{\Sigma}{2}\Tr\left[\mathcal{M}(\theta) U + \mathcal{M}^\dag(\theta)U^\dag\right] \, , \qquad \mathcal{M}(\theta) \equiv \ee^{-\ii \theta/2} \mathrm{diag}(m_u,m_d) \, ,
\eeq
where $U=\exp(\ii\pi_a\sigma_a/F_\pi)$ is the Nambu--Goldstone-boson field,
$F_\pi = \lim_{m_u,m_d\to 0} \frac{1}{\sqrt{2}m_\pi}\bra{0}
\overline{\psi}\gamma_0\gamma_5\psi \ket{\pi,\vec{p}=\vec{0}}$ is the pion
decay constant in the chiral limit, and $\Sigma = - \lim_{m_u,m_d\to
0}\braket{\overline{\psi}\psi}$ is the quark chiral condensate in the chiral limit.

At $T = 0$, the free energy density receives contribution only from the vacuum
energy, which can be estimated by minimizing the potential for the field $U$
appearing in $\mathcal{L}_{\eff}(\theta)$, i.~e., $V(\theta)=\Sigma \,
\Re\left[\mathcal{M}(\theta)U\right]$. That leads to the following
expression~\cite{DiVecchia:1980yfw} (see also
Refs.~\cite{GrillidiCortona:2015jxo,Luciano:2018pbj}):
\beq
F(T = 0, \theta) = - F_\pi^2 \left[ m_\pi^2(\theta) - m_\pi^2 \right] \ ; \ \ \ \ \ \ \ 
m_\pi^2 = \frac{\Sigma}{F_\pi^2}(1+z)m_u \ ; \ \ \ \ \ \ \ z=\frac{m_u}{m_d} \, .
\label{eq:f_chiPT}
\eeq
where $m_\pi (\theta)$ is the $\theta$-dependent pion mass:
\beq
m_\pi^2(\theta) = m_\pi^2\sqrt{1-\frac{4z}{1+z}\sin^2\left(\frac{\theta}{2}\right)} \, .
\label{eq:mpi_chiPT}
\eeq
Comparing this expression with the general parameterization in
Eq.~\eqref{eq:thetadep_free_energy_general} one finds, for the first two terms
in the $\theta$ expansion:
\beq
\chi = \frac{z}{1+z}m_\pi^2 F_\pi^2 = \Sigma \left(\frac{1}{m_u}+\frac{1}{m_d}\right)^{-1} \, , \qquad b_2 = -\frac{1}{12}\frac{1+z^3}{(1+z)^3} \, .
\label{eq:nf2_leading_chiPT}
\eeq
Therefore, as $m_u=m_d=m\to0$, the topological susceptibility vanishes linearly
with the quark mass, $\chi \sim m \to 0$, so that the theory becomes
$\theta$-independent as expected, while $b_2$, as well as the other $b_{2n}$
coefficients, tends to a constant.  It is interesting to notice that, for $m_u
= m_d$ ($z = 1$), this constant is smaller than the DIGA prediction by a factor
4, so the fact that $\chi \to 0$ cannot be simply interpreted in terms of a
more and more dilute gas of integer charged topological objects; however,
remarkably, the DIGA prediction is approached when $z \to 0$, i.~e. when $m_u
\to 0$ at fixed $m_d$, as can be seen explicitly also be Taylor expanding the
dependence of $m_\pi(\theta)$ in Eq.~(\ref{eq:mpi_chiPT}) and substituting it
in Eq.~(\ref{eq:f_chiPT}), which leads to $F(T,\theta) \propto (1 - \cos
\theta)$.

The computation above can be extended in several ways: for instance, the
strange quark contribution is easily added to the prediction for the
topological susceptibility, $\chi = \Sigma (1/m_u +1/m_d +1/m_s)^{-1}$.  A less
trivial extension is to go at the next-to-leading-order (NLO) of $\chi$PT
where, in the 2-flavour case, 10 additional low-energy constants appear:
details are not reported here, in the phenomenologically interesting cases
$z=1$ (corresponding to degenerate up and down quarks) and $z\simeq 0.48$
(corresponding to the physical up-down mass
ratio~\cite{FlavourLatticeAveragingGroupFLAG:2021npn}) one finally obtains the
following quantitative
predictions~\cite{Guo:2015oxa,GrillidiCortona:2015jxo,Bonati:2015vqz,Lu:2020rhp}:
\beq
\chi &= \left(75.5(5)~\mathrm{MeV}\right)^4, \quad b_2 &= -0.29(2), \quad z=0.48 \, ,\\
\chi &= \left(77.8(4)~\mathrm{MeV}\right)^4, \quad b_2 &= -0.22(1), \quad z=1 \, .
\eeq
For the topological susceptibility, also a N$^2$LO calculation is available in
the literature; however, the N$^2$LO contribution accidentally cancels with
$O(\alpha_{\rm em})$ LO corrections from Quantum Electrodynamics, thus
practically not altering these estimates~\cite{Gorghetto:2018ocs}.  Finally,
one can extend the analysis also to the finite temperature case, at least as
far as $T \lesssim T_{\rm pc}$: for this we refer the reader to
Ref.~\cite{GrillidiCortona:2015jxo}.

\subsection{The Lattice QCD approach: basics and results}

The purpose of this section is to provide a brief introduction to the
formulation of Lattice Gauge Theories and to the main issues and results
regarding the study of $\theta$-dependence by their means. Of course, any
reasonable level of detail about the general formulation is out of scope here,
for that we refer the reader to some standard textbooks on the
subject~\cite{Creutz:1983njd,Rothe:1992nt,Montvay:1994cy,Gattringer:2010zz}.

The most standard formulation is that on a hypercubic, isotropic lattice, with
lattice spacing $a$; gauge fields are represented by the elementary parallel
transporters connecting adjacent lattice sites, which are called {\em gauge
links} and belong to the gauge group, while quarks live on lattice sites.
Gluonic gauge invariant observables are written in terms of traces of gauge
link products over closed loops: the most elementary is the so-called {\em
plaquette}, which is related to a possible discretization of the pure gauge
action~\cite{Wilson:1974sk}. The covariant derivative in the quark action can
also be easily discretized in terms of gauge links, however some difficulties
are met to maintain at the same time locality, positivity, the correct number
of quark degrees of freedom, and a correct implementation of chiral
symmetry~\cite{Karsten:1980wd, Nielsen:1981hk}. The continuum limit is usually well defined: if
$N_{\subf}$ is not too large, asymptotic freedom guarantees the presence of a
critical point for a vanishing value of the bare gauge coupling $g$, where all
correlation lengths diverge and the system loses memory of discretization; the
dependence of the lattice spacing $a$ on $g$ is known by perturbation theory
for $g$ small enough, and can be determined non-perturbatively, i.~e., by
lattice simulations, otherwise.  If dynamical quarks are present, their masses
should be tuned with $g$, along the so-called {\em line of constant physics}
(LCP), in order to maintain physical observables (e.~g., the hadron masses)
unchanged as the continuum limit is approached.

Once discretized over a finite space-time volume, in order to deal with a
finite number of degrees of freedom, the partition function can be sampled by
Markov Chain Monte Carlo algorithms (if no sign problem is present), achieving
a statistical error on physical observables which decreases as the square root
of the statistics.  Systematic errors due to discretization can in principle be
completely removed by determining observables for several values of the lattice
spacings, then performing a continuum extrapolation. The needed computational
effort increases as $a$ decreases, both since the number of degrees of freedom
increases, and because the auto-correlation times of the sampling algorithms
grow as the continuum limit is approached ({\em critical slowing down}). The
presence of the fermion determinant, which is non-local in the gauge fields,
makes numerical simulations of full QCD much more demanding in terms of
computational power (even orders of magnitude) with respect to pure gauge
simulations, and this is especially true in the presence of light quarks and in
the low-temperature phase, because of the accumulation of small eigenvalues of
the Dirac operator due to chiral symmetry breaking~\cite{Banks:1979yr}. The
adoption of fermion discretizations which optimize their chiral properties
(Ginsparg--Wilson fermions~\cite{Ginsparg:1981bj}, in particular domain wall~\cite{Kaplan:1992bt,Shamir:1993zy,Furman:1994ky} or overlap~\cite{Neuberger:1997fp,Neuberger:1998wv,Luscher:1998pqa}) makes the needed effort much larger.\\

Topology is a property of continuum fields and of their homotopy classes,
therefore, strictly speaking, topology is not well defined on the lattice.
However, one expects the concept of topology to be recovered as the continuum
limit is approached, so that the lattice investigation of $\theta$-dependence
must rely on proper observables, capable of catching the emergence of topology,
with a well behaved continuum limit.

One possibility is to interpolate lattice fields by continuum ones, then
defining the winding number on the interpolation: that leads to the so-called
{\em geometric definitions} \cite{Berg:1981er, Luscher:1981zq, Woit:1983tq,
Phillips:1986qd}, which are typically non-polynomial functions of the lattice
fields. The lack of well defined homotopy classes then reflects in the presence
of lattice configurations with a large but finite action, typically associated
to large fluctuations at the lattice spacing scale, for which the interpolation
becomes ambiguous: such configurations, despite their large action, can in
principle proliferate as the lattice spacing is decreased, for entropic
reasons, thus obscuring the correct approach to the continuum limit
\cite{Berg:1981nw, Luscher:1981tq}. The alternative is to adopt a {\em
field-theoretic} discretization of $q(x)$, which is polynomial in the gauge
fields: however, in this case the discretized topological charge is non-integer
and renormalizations appear at various levels, e.~g., a multiplicative
renormalization \cite{Campostrini:1988cy,Campostrini:1989dh} for the
topological charge itself, and also additive renormalizations for the
susceptibility and the higher order cumulants
\cite{DiVecchia:1981aev,DiVecchia:1981hh, Campostrini:1989dh,DElia:2003zne},
with the latter which can be divergent and thus obscure the continuum limit.

One can also rely on the index theorem and look at the spectrum of a
discretized Dirac operator, in particular at its zero modes.  Standard lattice
discretizations of the Dirac operator do not possess exact zero modes with well
defined chirality: in this case one can go on and provide some definition based
on the spectrum and the corresponding eigenvectors, which is however subject to
renormalizations as well~\cite{Smit:1987zh,Smit:1987jh}.
On the other hand, when dealing
with lattice Dirac operators with exact zero-modes (like those satisfying the
Ginsparg--Wilson relation), one recovers an integer valued definition of the
winding number~\cite{Giusti:2001xh,Giusti:2004qd}, however ambiguities may emerge also in this case, see e.~g.
\cite{DelDebbio:2003rn,DelDebbio:2004ns}. 

A general, successful strategy, is to get rid of field fluctuations at the
lattice spacing scale, which are responsible for the renormalizations and
ambiguities mentioned above. For gluonic definitions, this is achieved by
smoothing techniques, like cooling \cite{Berg:1981nw, Iwasaki:1983bv,
Teper:1985rb, Ilgenfritz:1985dz, Campostrini:1989dh}, the gradient flow
\cite{Narayanan:2006rf,Luscher:2009eq, Luscher:2010iy, Luscher:2011bx}, or
other similar algorithms, all giving comparable
results~\cite{Bonati:2014tqa,Alexandrou:2017hqw}, which drive lattice
configurations obtained by Monte Carlo sampling towards minima of the
discretized action. For fermionic definitions, one achieves the goal of
removing additive renormalizations by projecting observables onto the space
generated by the eigenvectors of the lowest part of the Dirac spectrum, thus
removing contributions from the ultraviolet (UV) part of it
\cite{Luscher:2004fu,Giusti:2008vb, Luscher:2010ik, Bonanno:2019xhg}. The
strategy relies on the fact that the physical scale of topological fluctuations
becomes well separated, as the continuum is approached, from the UV scale of
the unphysical modes which are smoothed away; moreover, in the continuum
topology becomes stable under smoothing.

Topological properties of the gauge ensemble also affect the efficiency of
Monte Carlo sampling. A general phenomenon, affecting all theories with a
topological classification, is the so-called {\em topological freezing}: as the
continuum limit is approached, the emergence of infinite action barriers,
separating the different topological sectors, makes most Monte Carlo algorithms
unable to efficiently move from one sector to the other, thus leading to a
severe loss of ergodicity, usually much worse than standard critical slowing
down \cite{Alles:1996vn,DelDebbio:2002xa,DelDebbio:2004xh, Schaefer:2010hu, Bonati:2017woi}.

Further issues emerge for theories with light fermions.  In this case, the
fermion determinant should suppress the contribution from configurations with
non-zero winding number, because of the associated zero-modes of the Dirac
operator, thus leading to the expected suppression of $\theta$-dependence.
However, for discretized Dirac operators usually adopted for Monte Carlo
sampling, which do not have exact chiral properties, zero-modes are clearly
identified only quite close to the continuum limit, so that the observed
$\theta$-dependence can be misleading until sufficiently small lattice spacings
are reached, typically smaller than those needed for pure gauge theories. On
the other hand, the need to reach smaller lattice spacings makes the impact of
topological freezing more destructive.  \\

\begin{figure}[!t]
\centering
\includegraphics[scale=0.45]{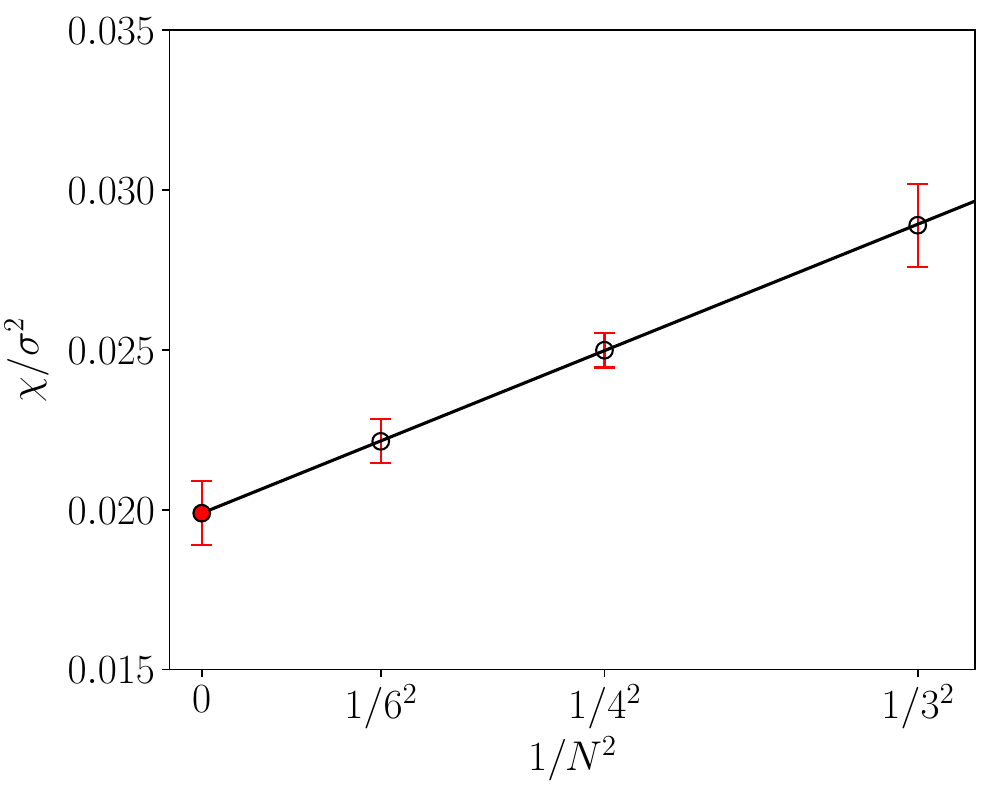}
\includegraphics[scale=0.45]{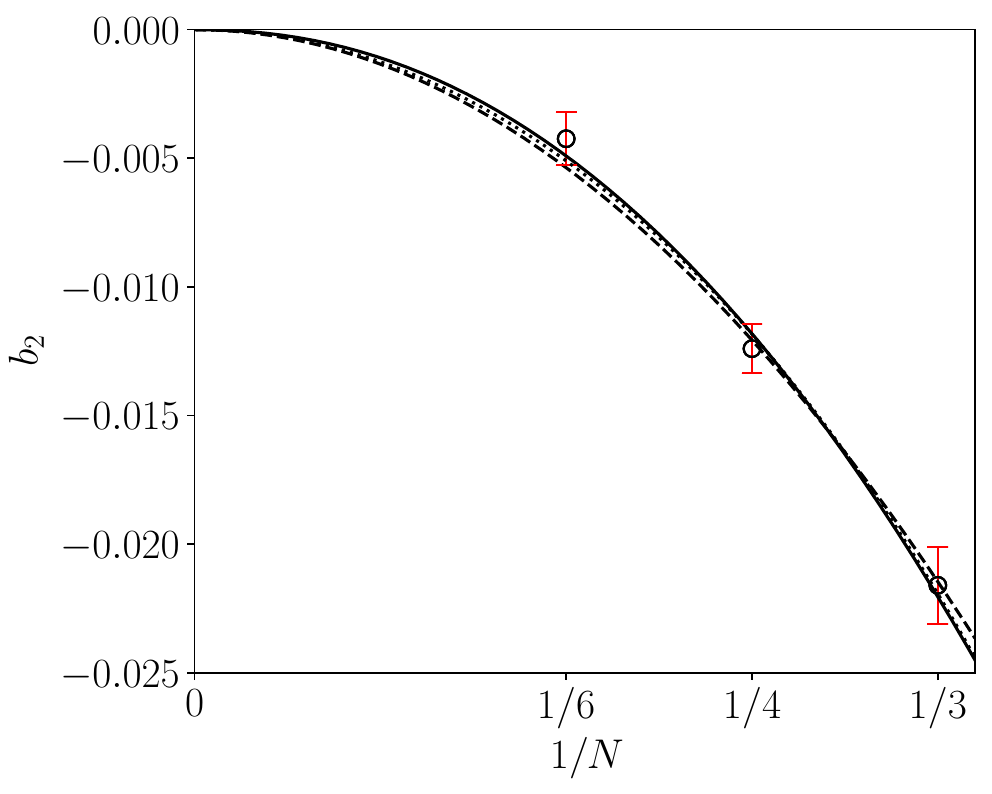}
\caption{Left: extrapolation of $\chi/\sigma^2$, where $\sigma$ is the string
tension, towards the large-$N$ limit, using the fit function
$\chi/\sigma^2=\bar{\chi}/\sigma^2+k/N^2$; the best fit yields
$\bar{\chi}/\sigma^2=0.0199(10)$ and $k=0.082(17)$. (source:
Ref.~\cite{Bonanno:2020hht}). Right: Extrapolation of $b_2$ towards the
large-$N$ limit (source: Ref.~\cite{Bonanno:2020hht}); a best fit according to
$b_2=\bar{b}_2/N^c$ with $c = 2$ yields  $\bar{b}_2 = - 0.193(10)$ (dashed
line), and is stable within errors by making $c$ a free parameter
($c=2.17(26)$, solid line) or by adding $1/N^4$ corrections (dotted line).
}
\label{fig:large_N_chi_b2}
\end{figure}

Let us now go to a brief review of available lattice results. The
Witten--Veneziano (WV) formula was the first available prediction about the
topological properties of Yang--Mills theories: it was based on large-$N$
considerations and phenomenological inputs, however its explicit confirmation
could only be based on some non-perturbative approach, based on first
principles.  Remarkably, the first pioneer lattice simulations of non-Abelian
gauge theories~\cite{Creutz:1979dw, Creutz:1980zw} were performed soon after
the proposal of the WV mechanism. Therefore, it is quite natural that the
topological susceptibility of SU($N$) pure gauge theories became soon one of
the most hunted prey for the first lattice pioneers.

The first attempts~\cite{DiVecchia:1981aev,DiVecchia:1981hh} were based on a
simple gluonic discretization of the topological charge density operator,
representing $F_{\mu\nu}$ in terms of an open plaquette, then subtracting an
additive renormalization for $\chi$ based on perturbative computations. Results
provided a non-zero value for $\chi$ in the SU(2) pure gauge theory, however
about 2 orders of magnitude smaller than predicted by the WV formula. Later
studies clarified that the problem was in a missing multiplicative
renormalization constant for the discretized topological charge
density~\cite{Campostrini:1988cy,Campostrini:1989dh}: after taking that into
account, later determinations led to results confirming the WV mechanism also
quantitatively, both for SU(2) and SU(3) pure gauge
theories~\cite{Alles:1996nm,Alles:1997qe}. The value of $\chi$ for pure gauge
SU(3) theory has been verified by many other studies, adopting gluonic
definitions in combination with smoothing techniques~\cite{Durr:2006ky, Ce:2015qha,
Bonati:2015sqt, Athenodorou:2020ani, Athenodorou:2021qvs, Bonanno:2023ple,
Durr:2025qtq}, or fermionic definitions~\cite{DelDebbio:2004ns, Luscher:2010ik,
Cichy:2015jra, Bonanno:2019xhg}, achieving an increasing control over the
continuum extrapolation and other possible systematic effects. Recently, also the so-called slope of the topological susceptibility, related to the 
leading-order momentum-dependence of the two-point
function of $q(x)$, has been determined~\cite{Bonanno:2023ple}.

The main point of the WV mechanism is however also that $\chi$ is finite in the
large-$N$ limit; this has been checked explicitly by various lattice
investigations \cite{Ce:2016awn, Bonati:2016tvi, Bonanno:2020hht,
Athenodorou:2021qvs}, which have determined $\chi$ for different values of $N$
and extrapolated it for $N\to\infty$: an example, taken from
Ref.~\cite{Bonanno:2020hht}, is reported in the left-hand side of
Fig.~\ref{fig:large_N_chi_b2}.  As discussed above, a finite value of $\chi$
for $N = \infty$ fixes, by large-$N$ arguments, the scaling with $N$ of the
higher-order coefficients $b_{2n}$ in the Taylor expansion of $F(T,\theta)$
around $\theta = 0$. As explicitly shown in Eq.~(\ref{eq:chib2n_def}), the
determination of such coefficients involves the determination of higher order
cumulants of the topological charge distribution at $\theta = 0$, which
represents a quite difficult numerical challenge: indeed, such cumulants
represent the deviations from a purely Gaussian behavior for the probability
distribution, however, by the central limit theorem, such deviations are less
and less visible as the total four-dimensional volume increases, so that a
significant increase in statistics is needed, in order to keep a constant
statistical error.

For this reason, early determinations of $b_2$ for the SU(3) pure gauge theory
were affected by statistical errors of the order of 20\% or
larger~\cite{DelDebbio:2002xa,DElia:2003zne,Giusti:2007tu}.  A substantial
improvement has been achieved by adding to the action an explicit source term
coupled to the topological charge, in particular a purely imaginary
$\theta$-term, so as to avoid the sign problem. In this way, the information
about the $b_{2n}$ coefficients is transferred to the dependence on the source
term of lower order cumulants, with a sizable reduction of statistical
errors~\cite{Panagopoulos:2011rb,Bonati:2015sqt}, bringing the uncertainty for
SU(3) well below the 10\% level. Further studies have looked at $b_2$ also for
larger values of $N$~\cite{Bonati:2016tvi,Bonanno:2020hht}, obtaining results
well compatible with the predicted $1/N^2$ scaling: an example, taken from
Ref.~\cite{Bonanno:2020hht}, is reported in the right-hand side of
Fig.~\ref{fig:large_N_chi_b2}.  For the same reason, the determination of the
other $b_{2n}$ coefficients turns out to be more and more challenging as $n$
increases: as a matter of fact, only $b_4$ has been determined, and
only for the SU(2) pure gauge theory~\cite{Bonanno:2018xtd}.

As discussed above, for SU($N$) gauge theories large-$N$ arguments can only
predict the scaling with $N$, however the measured values of the pre-factor can
teach something about the topological configurations dominating the
path-integral. For instance,  the fact that $b_2 \simeq - 0.19/N^2$ (see
Fig.~\ref{fig:large_N_chi_b2}) indicates that one expects at least sizable
corrections to the simple picture of a non-interacting gas of isolated objects
with fractional charge $1/N$, which would lead instead to $b_2 = -
0.08333/N^2$. This is consistent with the picture emerging from various lattice
investigations, which show that, in the low-temperature phase of Yang--Mills
theories, the topological charge density is typically delocalized in the
dominant gauge configurations (see, e.~g.,
Refs.~\cite{Buividovich:2011cv, Ilgenfritz:2007xu, Alexandru:2005bn}).

Till now we have discussed about lattice results regarding pure gauge theories.
In full QCD, including contributions from light dynamical fermions, $\chi$PT
provides quite solid and quantitative predictions for the $\theta$-dependence
of the zero temperature theory, like those reported in
Eq.~(\ref{eq:nf2_leading_chiPT}).  In this case, a lattice investigation of
$\theta$-dependence can be considered of moderate interest, as a check of the
validity of $\chi$PT and as an indirect probe of phenomenological low energy
constants such as $F_\pi$. What is maybe more interesting, is that it is a
stringent test of how well the lattice discretization is correctly implementing
the chiral properties of the theory, which should be visible, for instance,
from the linear scaling of $\chi$ with the quark mass (or with $m_\pi^2$): this
is a highly non-trivial issue, due to the well known difficulties in
formulating chiral fermions on a lattice.

In the framework of a numerical computation of the QCD path-integral, dynamical
fermions are encoded in the fermion determinant, which for light masses will
act as a suppression term for the statistical weight of gauge configurations
with zero-modes of the Dirac operator, i.~e. with non-trivial topology.
Therefore, a failure of the discretized Dirac operator to correctly identify
such zero-modes, will result in a incorrect distribution of the topological
charge. While the commonly used lattice discretizations are expected to
eventually recover the correct chiral properties close enough to the continuum
limit, this effect will typically add to other systematics related to
discretization, making the approach to the continuum limit slower.
For this reason, as well as for the intrinsic efforts required by simulations
with dynamical fermions, reliable investigations of $\theta$-dependence at the
physical point, or approaching the chiral limit, have been available only
recently; by now, several consistent studies exist \cite{Bonati:2015vqz,
Borsanyi:2016ksw,Alexandrou:2017bzk,Athenodorou:2022aay,Bonanno:2023xkg}.  

\begin{figure}[!t]
\centering
\includegraphics[scale=0.48]{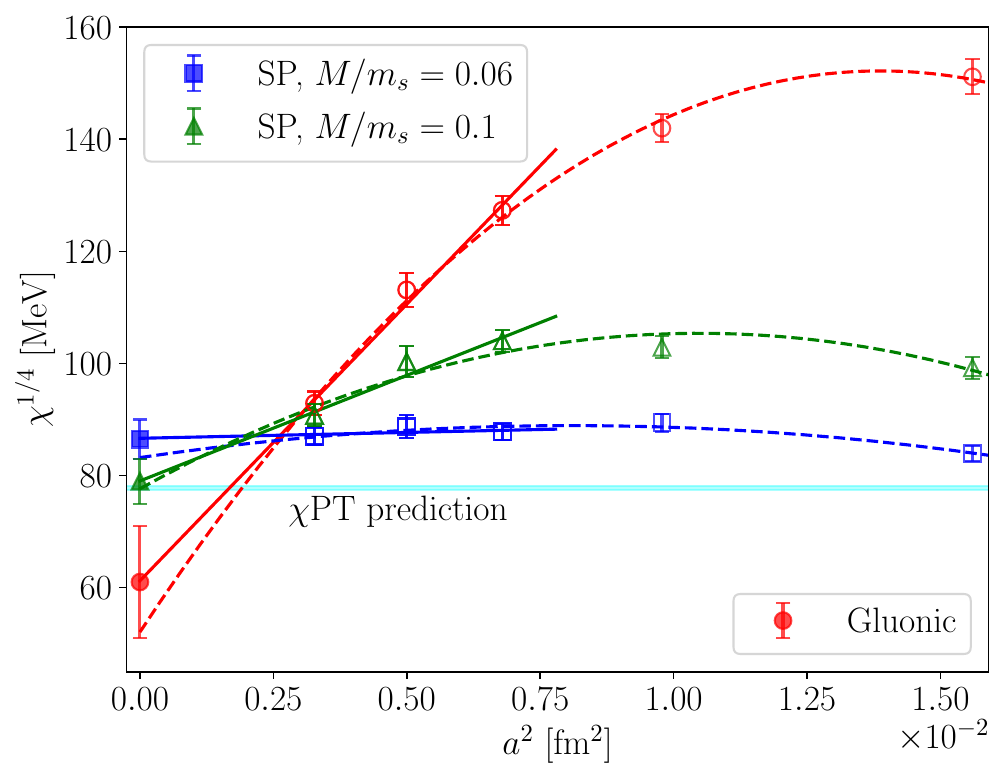}
\includegraphics[scale=0.43]{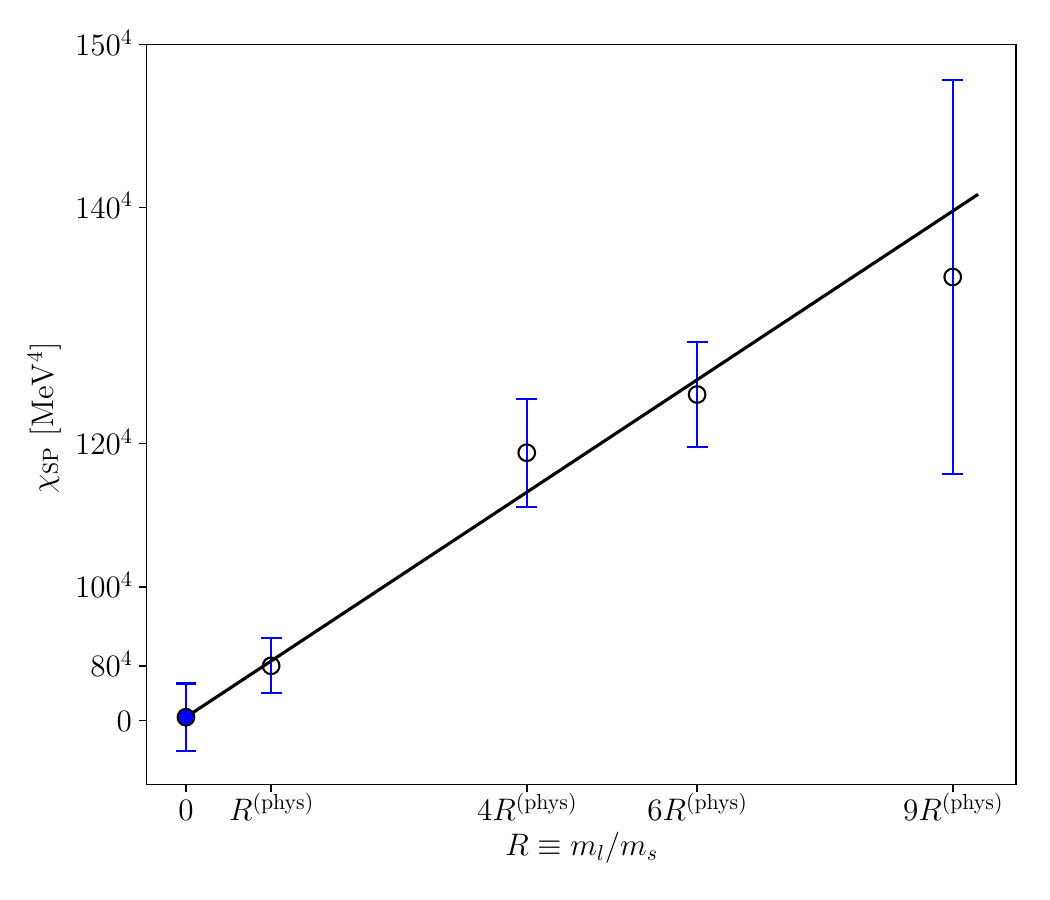}
\caption{Left: continuum extrapolation of $\chi^{1/4}$ at $T = 0$, for $N_{\subf} =
2+1$ QCD at the physical point (discretized via staggered fermions) and various
definitions of topological observables; SP stands for Spectral Projectors, with
$M/m_s$ being the cutoff on the Dirac spectrum used for the projection; figure
adapted from Ref.~\cite{Athenodorou:2022aay}. Right: for the same theory and
discretization, continuum extrapolated values of $\chi$ for different values of
the light-to-strange quark mass ratio, and extrapolation to the chiral limit;
figure taken from Ref.~\cite{Bonanno:2023xkg}.}
\label{fig:continuum_full_gluonic_spectral}
\end{figure}

As an example, in left-hand side of
Fig.~\ref{fig:continuum_full_gluonic_spectral}, which is taken from
Ref.~\cite{Athenodorou:2022aay}, we show $\chi^{1/4}$ as a function of the
lattice spacing, determined through various approaches (either gluonic or
fermionic) for QCD with $N_{\subf} = 2+1$ and physical quark masses: as it is
clearly visible, for lattice spacings for which discretization errors are
usually well under control, e.~g., $a = 0.1$~fm, the value of $\chi$ is still
more than one order of magnitude larger than the $\chi$PT expectation when
using the standard gluonic definition of winding number; only after the
continuum extrapolation, the various determinations converge to the $\chi$PT
prediction. In the right-hand side of the same figure, which is taken from
Ref.~\cite{Bonanno:2023xkg}, we report the continuum extrapolated values of
$\chi$ for $N_{\subf} = 2+1$ QCD and different values of the light quark mass,
which show the correct approach to the chiral limit predicted by $\chi$PT.  The
need to reach lattice spacings much smaller than usual, makes the impact of
topological freezing more significant; for this reason, various possible
approaches to mitigate the problem are under investigation in simpler (often
lower-dimensional) models, or are being applied to the case of full
QCD~\cite{Vicari:1992jy, Brower:2003yx, Aoki:2007ka, Luscher:2011kk,
LSD:2014yyp, Mages:2015scv, Bietenholz:2015rsa, Laio:2015era, Bonati:2017woi,
Hasenbusch:2017unr, Luscher:2017cjh, Bonanno:2020hht, Florio:2019nte,
Funcke:2019zna, Kanwar:2020xzo, Nicoli:2020njz, Albandea:2021lvl,
Cossu:2021bgn, Borsanyi:2021gqg, Fritzsch:2021klm,
Abbott:2023thq, Eichhorn:2023uge, Howarth:2023bwk, Albandea:2024fui,
Bonanno:2024udh, Vadacchino:2024lob, Abe:2024fpt, Bonanno:2024zyn}.

\subsection{High temperature: \texorpdfstring{$\theta$}{theta}-dependence across the phase transition}
\label{sec:HT}

As we have discussed above, one expects that the DIGA should give the correct
description of $\theta$-dependence at least for asymptotically high
temperatures, because of the natural infrared cut-off on large size instantons;
however, how high should the temperature be for the DIGA regime to sets in, is
in principle not known. On the other hand, we know that in the large-$N$ limit
DIGA predicts $\chi = 0$; therefore, if $\chi \neq 0$ in the same limit in the
zero-$T$ regime, as predicted by the WV mechanism and verified by lattice
simulations, there must be a non-analytic point between the two regimes,
i.~e.~a phase transition, at least in the $N \to \infty$ limit. It is then a
quite natural expectation that this transition should coincide with the only
other finite $T$ phase transition which is known for SU($N$) gauge theories in
the large-$N$ limit, namely the deconfinement transition. One can also further
conjecture that a change in $\theta$-dependence should
persist also for finite values of $N$, still taking place around the
deconfinement transition. Remarkably, this is exactly what has been
observed by lattice simulations, as we briefly review in the following.

Early indications for a suppression of $\chi$ across the deconfinement
transition of the SU(2) pure gauge theory~\cite{Teper:1985gi,Teper:1985ek} were
later confirmed both for SU(3)~\cite{Alles:1996nm} and
SU(2)~\cite{Alles:1997qe}. In particular, the topological susceptibility
appears to be practically independent of $T$ as long as $T < T_{\rm c}$, where $T_{\rm c}$
is the deconfinement transition, then rapidly dropping to small but non-zero
values for $T > T_{\rm c}$.  The extension of such results to larger values of
$N$~\cite{DelDebbio:2004vxo} then showed that $\chi$ seems indeed to vanish, as
$N \to \infty$, for any $T > T_{\rm c}$, as expected from the argument above. Recently, this has been 
further corroborated by a calculation of the susceptibility at the critical temperature in the 
deconfined phase $\chi_{\rm d}^{+} = \chi(T=T_{\rm c}^{+})$~\cite{Borsanyi:2022fub,Bonanno:2023hhp}, 
showing that this quantity vanishes as $\sim\exp(-N)$ approaching the large-$N$ limit~\cite{Bonanno:2023hhp}. This is a
strong indication that the DIGA regime indeed sets in, also for finite $N$, as
soon as the system crosses deconfinement: further studies of pure gauge
topology at finite $T$ have further refined this
picture~\cite{Berkowitz:2015aua, Borsanyi:2015cka, Frison:2016vuc,
Kitano:2015fla,Borsanyi:2021gqg} looking also for
the asymptotic behavior of $\chi$ as a function of $T$ predicted at the 1-loop
level, see Eq.~(\ref{eq:asym_chi_DIGA}), which, as we have stressed above,
could set in at temperatures much larger than those at which the diluteness
hypothesis starts to be valid.

Actually, the suppression of $\chi$, even if strongly suggestive, cannot be
considered by itself as conclusive evidence for the validity of DIGA, i.~e. of
the hypothesis that the dominant topological contributions to the functional
integral can be described as the superposition of a diluted ensemble of
non-interacting objects carrying integer topological charge. Indeed, such
hypothesis leads to a precise prediction for $F(T,\theta) \propto (1 - \cos
\theta)$, see Eq.~(\ref{eq:free_energy_theta_DIGA}), which implies the values
of the $b_{2n}$ coefficients reported in Eq.~(\ref{eq:b2n_DIGA}), e.~g., $b_2 =
-1/12 \simeq -0.833$.  There are some counterexamples where a vanishing $\chi$
does not correspond to that, like the chiral limit of QCD with non-vanishing
$m_u / m_d$ (see our discussion after Eq.~(\ref{eq:nf2_leading_chiPT}), or
two-dimensional $\CP^{N-1}$ models, for which in the $N \to \infty$ limit
$\chi$ vanishes as $1/N$~\cite{DAdda:1978vbw, Luscher:1978rn, Witten:1978bc,
Campostrini:1991kv, Campostrini:1991tw}, however the $b_{2n}$ coefficients
vanish as well \cite{DelDebbio:2006yuf, Rossi:2016uce, Bonati:2016tvi,
Sugeno:2025exv}.

For this reason, the lattice investigation of $b_2$ as a function of $T$
reported in Ref.~\cite{Bonati:2013tt} represented an important consistency
check of the change of $\theta$-dependence happening across the deconfining
transition, proving that indeed $b_2$ goes to the DIGA prediction soon after
$T_{\rm c}$, and increasingly faster as $N$ increases.  Such results have been later
confirmed by other studies~\cite{Borsanyi:2015cka, Vig:2021oyt, Kovacs:2023vzi, Yamada:2024vsk}
and extended to other theories characterized by a deconfining transition, like
the 4D gauge theory based on the G$_2$ gauge group~\cite{Bonati:2015uga}, thus
proving a general link between the change of $\theta$-dependence and
deconfinement.

When considering lattice investigations of $b_2$ above $T_{\rm c}$, one should
actually be careful about the following possible caveat.  Since $\chi$ drops
rapidly above $T_{\rm c}$, and since lattice simulations are always constrained to
finite volumes, it may easily happen that at high enough temperatures (or too
small volumes) $\langle Q^2 \rangle = \mathcal{V}\, \chi \ll 1$. In this
situation,  the gauge ensemble is essentially composed by configurations with
$Q = 0$, with only a small fraction $\epsilon$ with $|Q| = 1$ and practically
none with $|Q| > 1$, so that $Z(T,\theta) \propto (1 + \epsilon \cos \theta)$
and $F(T,\theta) \propto (1 - \cos \theta)$.  However, this is clearly a fake
DIGA behaviour, driven by finite volume effects, which is quite different from
a fair DIGA behaviour in which $\langle Q^2 \rangle \gtrsim 1$ and nevertheless
one has $F(T,\theta) \propto (1 - \cos \theta)$ because of the Poissonian
distribution of integer-charged topological excitations. The lattice
determinations of $b_2$ across $T_{\rm c}$ are typically not affected by this
possible caveat.

Interestingly, the change of $\theta$-dependence across the phase transition
drives a corresponding dependence of the critical temperature $T_{\rm c}$ on
$\theta$~\cite{Sasaki:2011cj,Unsal:2012zj,DElia:2012pvq,Otake:2022bcq}.  Indeed, $T_{\rm c}$ is
fixed by the exact balance between the free energies of the two phases,
confined and deconfined, so that, if the dependence of the two free energies on
$\theta$ is different, the balance point is expected to change with $\theta$;
in particular, since the confined free energy grows more rapidly (because of
the large value of $\chi$), the values of $T_{\rm c}$ is expected to drop with
$\theta$, with a leading behaviour proportional to
$\theta^2/N^2$~\cite{DElia:2012pvq}. This has been explicitly verified by
lattice simulations performed at imaginary values of $\theta$, first for $N =
3$~\cite{DElia:2012pvq,DElia:2013uaf}, then also for $N =
2$~\cite{Yamada:2024pjy} and $N > 3$~\cite{Bonanno:2023hhp}.

The introduction of dynamical fermions does not change the picture
significantly.  One important point is that, in QCD with physical or
close-to-physical quark masses, the deconfinement phase transition is
substituted by a rapid crossover at a pseudocritical temperature $T_{\rm pc}$,
at which the (approximate) chiral symmetry gets restored, with $T_{\rm pc}
\simeq 155$~MeV for physical quark masses. Hence, one can in principle expect
that also the change in $\theta$-dependence could be spread over a wider range
of temperatures, with respect to the pure gauge case.  The drop of the
topological susceptibility across the crossover has been typically confirmed by
various lattice studies over the years~\cite{Alles:2000cg, Bonati:2015vqz,
Borsanyi:2016ksw, Petreczky:2016vrs, Bonati:2018blm, Burger:2017xkz,
Athenodorou:2022aay,Chen:2022fid, Kotov:2025ilm} and proved to extend also to
other transitions in the QCD phase diagram, like 
in the presence of background magnetic
fields~\cite{Brandt:2024gso} or for two-color QCD at finite
density~\cite{Alles:2006ea,Hands:2011hd,Iida:2019rah,Lombardo:2021jpn,Kawaguchi:2023olk,Iida:2024irv,Itou:2025vcy}, with
the possible exception of the low-$T$ and high density region of the phase diagram~\cite{Iida:2019rah,Iida:2024irv,Itou:2025vcy}.
The interesting questions are whether the drop
of $\chi$ corresponds also in this case to the onset of the DIGA regime, and
what is the exact behavior of $\chi(T)$ for temperatures substantially larger
than $T_{\rm pc}$; the latter question is particularly important for axion
cosmology, since various models permit to fix the axion coupling constant
$f_a$, hence its mass, based on $\chi(T)$ and on the observed dark matter
abundance \cite{Preskill:1982cy, Abbott:1982af, Dine:1982ah, Wantz:2009it,
DiLuzio:2020wdo}.  Interesting information about axion physics, in particular
regarding the axion thermal production~\cite{Notari:2022ffe}, can be obtained
also from the Euclidean time correlator of the topological charge density,
which permits to determine the so-called {\em sphaleron
rate}~\cite{Kotov:2018vyl, Altenkort:2020axj, BarrosoMancha:2022mbj,
Bonanno:2023ljc, Bonanno:2023thi}.

The determination of $b_2$ has shown a clear approach to the DIGA prediction
above $T_{\rm pc}$, even if possibly slower than in the pure gauge
case~\cite{Bonati:2015vqz,Kotov:2025ilm}, and also the drop of
$\chi(T)$ is compatible with the 1-loop DIGA prediction in
Eq.~(\ref{eq:asym_chi_DIGA}), at least for the power-law exponent, already
starting from unexpectedly low values of
$T$~\cite{Borsanyi:2016ksw,Petreczky:2016vrs,Bonati:2018blm,
Athenodorou:2022aay,Kotov:2025ilm}. The determination of $\chi(T)$ turns
out to be a particularly challenging problem, since various numerical problems
meet at the same time: discretization effects induced by the bad chiral
properties of the fermion determinant can lead to misleading results (see,
e.~g., the early results reported in Ref.~\cite{Bonati:2015vqz}); the need to
reach quite small lattice spacings exacerbates problems related to topological
freezing; finally, the very small values reached by $\chi$ at high $T$ mean
that, even on reasonable volumes, one has $\langle Q^2 \rangle = \mathcal{V}\,
\chi \ll 1$, which makes the statistical accuracy on $\chi$ quite low, in this
case one can rely on multicanonical approaches to partially mitigate this
problem~\cite{Bonati:2017woi, Bonati:2018blm, Athenodorou:2022aay}.  \\ \\

%% file: sec4_concl.tex
\section{Conclusions}

In this chapter we introduced the $\theta$-dependence of four-dimensional gauge
theories and some of its main features. In particular, we discussed several
approximation schemes that can be used to get quantitative or, in some cases,
semi-quantitative information on the $\theta$-dependence of QCD and QCD-like
theories in various regimes.  We then presented a short review of recent
results obtained by carrying out numerical simulations of the lattice
discretized theory.  

While $\theta$-dependence provides fundamental insights in several aspects of
the low-energy QCD phenomenology, and especially for what concerns the
pseudoscalar meson spectrum, it also opens new problems. The most famous of
these problems is the strong CP problem, which is related to the experimental
absence of CP violations in the strong sector of the Standard Model, despite
the fact that the $\theta$-term explicitly breaks CP.  This fact forces the
physical value of the $\theta$ angle to be surprisingly small,
$|\theta|\lesssim 10^{-10}$. 

The axion solution of the strong CP problem is definitely the most investigated
one, mainly because of the phenomenological appeal of the axion as a dark
matter candidate. Since several axion properties (in particular its mass and
self-coupling) can be directly related to $\theta$-dependence, in the last
years there has been a renewed interest in the $\theta$-dependence of QCD at
high temperature. It is indeed possible to use the high-temperature behavior of
the topological susceptibility $\chi(T)$ (more precisely: $\chi(T)$ in the
range of temperatures between $\approx 150$~MeV and $O(10)$~GeV) to put an upper
bound on the axion decay constant.
Despite the enormous progress made in recent years, several important questions
regarding $\theta$-dependence remain open or at least need to be explored
further. 

The most important short-term goal is clearly to reach a quantitative consensus
regarding the values of $\chi(T)$ for temperatures up to at least a few GeV. As
discussed in Sec.~\ref{sec:HT} a coherent global picture of $\theta$-dependence
in the high temperature regime of QCD in by now well established, however
quantitative (and in a few rare cases even qualitative) discrepancies exist
between the results obtained by different groups, using different lattice
discretizations and analysis techniques. Concerning this point it is important
to stress once more that computations of $\chi(T)$ at high temperature, carried
out using physical light quark masses, are extremely challenging. To carry out
QCD computation with physical light quark masses is already a major
computational challenge  in standard conditions, but topology make things even
worst. Different fermionic lattice discretizations indeed present different
lattice artifacts, and results obtained by using different approaches have to
be properly extrapolated to the continuum limit to be compared with each other.
Topological freezing and the slow continuum convergence of the low-laying Dirac
spectrum make however very difficult to perform controlled continuum limit
extrapolations, and this is likely the main reason for the discrepancies
present in the literature. To improve this state of affairs the development of
new sampling algorithms, which will allow us to perform simulations at smaller
lattice spacings, is of the paramount importance, and indeed this is a very
active research area, in which some traditional and well-established techniques
are being applied together with more innovative approaches, see, e.~g.,
\cite{Vicari:1992jy, Brower:2003yx, Aoki:2007ka, Luscher:2011kk, LSD:2014yyp,
Mages:2015scv, Bietenholz:2015rsa, Laio:2015era, Bonati:2017woi,
Hasenbusch:2017unr, Luscher:2017cjh, Bonanno:2020hht, Florio:2019nte,
Funcke:2019zna, Kanwar:2020xzo, Nicoli:2020njz, Albandea:2021lvl,
Cossu:2021bgn, Borsanyi:2021gqg, Fritzsch:2021klm, Abbott:2023thq,
Eichhorn:2023uge, Howarth:2023bwk, Albandea:2024fui, Bonanno:2024udh,
Vadacchino:2024lob, Abe:2024fpt, Bonanno:2024zyn}.

Concerning the systematics of numerical simulations, it is also worth noting that
an extrapolation from unphysical (larger than physical) to physical light quark
masses was often used in the past to reduce the computational burden of zero
temperature QCD simulations. In studies of $\chi(T)$ at finite temperature this
strategy presents however several difficulties, since the determination of the
functional form to be used in the mass extrapolation is a fundamental part of
the problem under investigation. At low-temperature we can safely rely on
$\chi$PT, at very high-temperature one expects semiclassical DIGA predictions
to be reliable, but for temperatures close to the chiral crossover, or anyway
of the order of a few GeV, no theoretically solid framework is available to
guide the extrapolation. The precise mass dependence of the finite-temperature
Dirac spectrum is also related to the possible effective restoration of the
U(1)$_{\subA}$ symmetry at finite temperature; this topic has been studied by
several research groups~\cite{Edwards:1999zm, Dick:2015twa, Aoki:2020noz,
Ding:2020xlj, Aoki:2021qws, Kaczmarek:2021ser, Kaczmarek:2023bxb,
Kovacs:2023vzi, Azcoiti:2023xvu, Giordano:2024jnc, Alexandru:2024tel,
Fodor:2025yuj} but no general consensus has been reached yet.

When discussing lattice result in Sec.~\ref{sec3} we focused on $\theta$-dependence for small values of $\theta$ and at equilibrium, discussing aspects
that are, in practice, related to the second and to the fourth derivative of the
free energy density $f(T,\theta)$ at $\theta=0$. In this regime, although some
delicate points still need to be further investigated, we have an overall rather
clear physical picture. Concerning what happens for larger values of $\theta$,
and in particular at $\theta=\pi$, there are instead only exploratory
studies~\cite{Kitano:2017jng, Kitano:2020mfk, Hirasawa:2024fjt} with more
reliable results existing only for systems in reduced
dimensionality~\cite{Alles:2007br, Alles:2014tta, Sulejmanpasic:2020lyq,
Sugeno:2025exv, Matsumoto:2025fjb}. Direct simulations at real $\theta$ values
are prevented by the sign problem, and for large $\theta$ it is difficult to
obtain reliable results from imaginary-$\theta$ simulations or Taylor expansion
at $\theta=0$, due to the difficulties of performing the analytic continuation
or to evaluate a large enough number of coefficients of the expansion. Real time
results, relevant to describe out of equilibrium processes, also present huge
difficulties, mainly (but not only) related to the resolution of the inverse
problem required to extract spectral densities. The only results available so
far in this context are those regarding the sphaleron rate \cite{Kotov:2018vyl,
Altenkort:2020axj, BarrosoMancha:2022mbj, Bonanno:2023ljc, Bonanno:2023thi}.